\documentclass[fleqn,usenatbib]{mnras}
\usepackage{newtxtext,newtxmath}

\usepackage[T1]{fontenc}
\usepackage{ae,aecompl}

\usepackage{graphicx}	

\title[Hydrogen Recombination Laser Lines in Mz~3]{Herschel Planetary Nebula Survey
(HerPlaNS)\thanks{{\it Herschel} is an ESA space observatory with science instruments provided
by European-led Principal Investigator consortia and with important participation from NASA.}: Hydrogen Recombination Laser Lines in Mz~3}
\author[I. Aleman et al.]{Isabel Aleman$^{1,2}$\thanks{E-mail: isabel.aleman@usp.br},
Katrina Exter$^{3,4,5}$,
Toshiya Ueta$^{6}$,
Samuel Walton$^{2,7}$,
\newauthor{A. G. G. M. Tielens$^{2}$,
Albert Zijlstra$^{8,9}$,
Rodolfo Montez Jr.$^{10}$,
Zulema Abraham$^{1}$,
}
\newauthor{Masaaki Otsuka$^{11}$,
Pedro P. B. Beaklini$^{1}$,
Peter A. M. van Hoof$^{12}$,
Eva Villaver$^{13}$,
}
\newauthor{Marcelo L. Leal-Ferreira$^{14}$,
Edgar Mendoza$^{1}$,
Jacques D. R. L\'{e}pine$^{1}$
}
\\
$^{1}$ IAG-USP, University of S\~{a}o Paulo, Rua do Mat\~{a}o 1226, Cidade Universit\'{a}ria, 05508-090, S\~{a}o Paulo, SP, Brazil\\
$^{2}$ Leiden Observatory, University of Leiden, PO Box 9513, 2300 RA, Leiden, The Netherlands\\
$^{3}$ Institute of Astronomy, KU Leuven, Celestijnenlaan 200D, BUS 2401, 3001 Leuven\\
$^{4}$ Herschel Science Centre, European Space Astronomy Centre, ESA, P.O.Box 78, Villanueva de la Ca\~{n}ada, Spain\\
$^{5}$ ISDEFE, Beatriz de Bobadilla 3, 28040 Madrid, Spain\\
$^{6}$ Department of Physics and Astronomy, University of Denver, 2112 E. Wesley Ave., Denver, CO 80210, USA\\
$^{7}$ Astrophysics Research Institute, Liverpool John Moores University, IC2, Liverpool Science Park, 146 Brownlow Hill, Liverpool L3 5RF, UK\\
$^{8}$ Jodrell Bank Centre for Astrophysics, Alan Turing Building, University of Manchester, Manchester, M13 9PL, UK\\
$^{9}$ Department of Physics \& Laboratory for Space Research, University of Hong Kong, Pok Fu Lam Rd., Hong Kong\\
$^{10}$ Smithsonian Astrophysical Observatory, Cambridge, MA 02138, USA\\
$^{11}$ Institute of Astronomy and Astrophysics, 11F of Astronomy-Mathematics Building, AS/NTU. No.1, Section 4, Roosevelt Rd., \\Taipei 10617, Taiwan, ROC\\
$^{12}$ Royal Observatory of Belgium, Ringlaan 3, B-1180, Brussels, Belgium\\
$^{13}$ Departamento de F\'{i}sica Te\'{o}rica, Universidad Aut\'{o}noma de Madrid, Cantoblanco, E-28049, Madrid, Spain\\
$^{14}$ Argelander-Institut f\"ur Astronomie, Universit\"at Bonn, Auf dem H\"ugel 71, 53121 Bonn, Germany\\
}

\date{Accepted XXX. Received YYY; in original form ZZZ}
\pubyear{2015}
\begin{document}
\label{firstpage}
\pagerange{\pageref{firstpage}--\pageref{lastpage}}
\maketitle

\begin{abstract}
The bipolar nebula Menzel~3 (Mz~3) was observed as part of the \textit{Herschel} Planetary Nebula Survey (\textit{HerPlaNS}), which used the PACS and SPIRE instruments aboard the \textit{Herschel Space Observatory} to study a sample of planetary nebulae (PNe). In this paper, one of the series describing \textit{HerPlaNS} results, we report the detection of H~{\sc i}~recombination lines (HRLs) in the spectrum of Mz~3. Inspection of the spectrum reveals the presence of 12 HRLs in the 55 to 680~$\mu$m range covered by the PACS and SPIRE instruments (H11$\alpha$ to H21$\alpha$ and H14$\beta$). The presence of HRLs in this range is unusual for PNe and has not been reported in Mz~3 before. Our analysis indicates that the HRLs we observed are enhanced by laser effect occurring in the core of Mz~3. Our arguments for this are: (i) the available Mz~3 optical to submillimetre HRL $\alpha$ line intensity ratios are not well reproduced by the spontaneous emission of optically thin ionized gas, as would be typical for nebular gas in PNe; (ii) the compact core of Mz~3 is responsible for a large fraction of the \textit{Herschel} HRLs emission; (iii) the line intensity ratios for Mz~3 are very similar to those in the core emission of the well known star MWC~349A, where laser effect is responsible for the enhancement of HRLs in the \textit{Herschel} wavelength range; (iv) the physical characteristics relevant to cause laser effect in the core of MWC~349A are very similar to those in the core of Mz~3. 
\end{abstract}

\begin{keywords}
                 planetary nebulae: general --
                 planetary nebulae: individual: Mz 3 --
                 binaries: symbiotic --
                 stars: emission-line, Be --
                 circumstellar matter --
                 masers
\end{keywords}


\section{Introduction} \label{intro}

\setlength{\skip\footins}{0.4cm}

Menzel 3 (Mz~3; Ant Nebula; PN 331.7-01.0) is a puzzling object. It has been classified as a planetary nebula (PN), a pre-PN, a symbiotic star, and a symbiotic Mira, but its true nature is still not understood \citep[e.g.,][]{1978ApJ...221..151C, 1983MNRAS.204..203L, 2001A&A...377L..18S, 2004MNRAS.354..549B,2011MNRAS.413..514C}. \citet{2010A&A...509A..41C} placed Mz~3 on the borderline between a symbiotic star and a young PNe, since it has characteristics from both classes of objects.

Mz~3 has a clear bipolar morphology, with a very narrow waist and symmetrically opposed lobes \citep{1978ApJ...221..151C}. The nebula exhibits multiple outflows and a number of substructures, including a puzzling equatorial ring \citep{1985MNRAS.215..761M,2004AJ....128.1694G,2004A&A...426..185S,2015A&A...582A..60C}. The outflows along the polar axis can reach hypersonic velocities \citep[$\sim$500~km~s$^{-1}$;][]{1985MNRAS.215..761M, 2000MNRAS.312L..23R, 2004A&A...426..185S}. In the core of the nebula, a disc is obscuring the central source \citep[e.g. ][]{1978ApJ...221..151C, 2003ApJ...591L..37K, 2004MNRAS.354..549B}. \citet{2007A&A...473L..29C} characterised this disc as rather flat, seen nearly edge-on and rich in amorphous silicates. \textit{Spitzer Space Telescope IRAC} colours indicate the presence of warm dust at the core of Mz~3 \citep{2011MNRAS.413..514C}. The density in the core exceeds 10$^6$~cm$^{-3}$ \citep{2003MNRAS.342..383S, 2002MNRAS.337..499Z}. 
\citet{2005A&A...444..861P} found that the lobes are composed of ionized gas with $n_{\rm H} \sim$ 4 $\times$ 10$^3$~cm$^{-3}$, probably ejected in multiple events, and that the ionizing star has $T_{\rm eff} =$~39\,300~K and $L_{\star} =$~9100~$L_{\sun}$, which are similar to the values obtained by \citet{2003MNRAS.342..383S} (30\,000~K and 10\,000~L$_{\sun}$, respectively) using a different procedure. \citet{2003ApJ...591L..37K} detected bright X-ray emission from the core and a possible jet. Mz~3 is also a very bright radio emitter \citep{2004MNRAS.354..549B,2007ApJ...665..341L}. The distance to Mz~3 has been estimated to values between 1.0 and 2.7~kpc \citep[e.g.][]{1978ApJ...221..151C,1983MNRAS.204..203L,1992MNRAS.259..635K,2003MNRAS.342..383S,2005A&A...444..861P}

Here, we add a new interesting feature to the list of Mz~3 characteristics by reporting the detection of H~{\sc i}~ recombination lines (HRLs) in its \textit{Herschel} far-infrared (FIR) to submillimetre (submm) spectrum, which we propose is enhanced by laser effect produced in the dense core of this nebula\footnote{It is interesting to note that D. H. Menzel was one of the first scientists to suggest that negative opacities (i.e. light amplification by stimulated emission, later named laser effect) might occur in certain conditions \citep{1937ApJ....85..330M} and we now detect laser emission in one of the PNe he discovered \citep{1922BHarO.777....0M}.}. The study of HRL lasers\footnote{For simplification, in the remainder of the text we will use laser to designate the amplification by stimulated emission in all ranges of the electromagnetic spectrum.} is an important tool for inferring physical conditions and kinematics in compact ionized regions, where typical diagnostics, such as forbidden line ratios, are suppressed. As discussed, for example by \citet{2000A&A...361.1169H} and \citet{1996ApJ...470.1118S}, laser effect on HRLs occurs in a somewhat narrow range of physical conditions. Models for the profiles of such lines can provide a detailed view of the physical structure and kinematics, as demonstrated in the studies of \citet{1996ApJ...470.1134S}, \citet{2011A&A...530L..15M}, and \citet{2013A&A...553A..45B} of the unresolved core of the B[e] star MWC~349. A detailed discussion of the laser effect on HRL lines and its applications can be found in \citet{1996ApJ...470.1118S}.

This paper is organized as follows: Sect.~\ref{observ} \S{2} describes the observations and the data reduction method; Sect.~\ref{firspec} presents an overview of the Mz~3 FIR/submm spectral features; Sect.~\ref{recomb} presents the detection of FIR/submm HRLs and their characteristics; Sect.~\ref{discuss} discusses the probable laser nature of the hydrogen lines; conclusions are summarized in Sect.~\ref{conclude}.

\section{Observations} \label{observ}

\begin{figure}
   \centering
   \includegraphics[width=8.2cm]{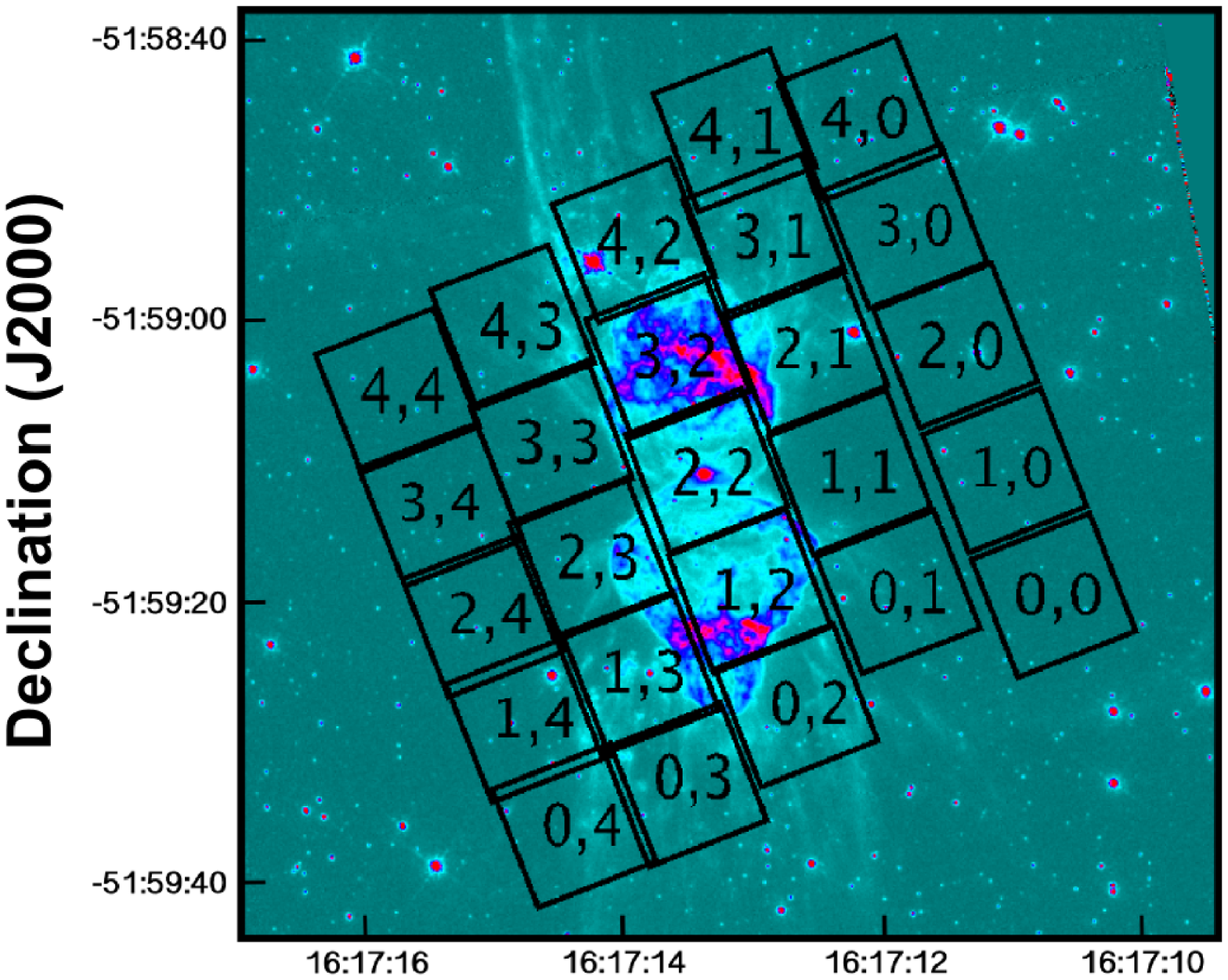}\\
   \includegraphics[width=8.2cm]{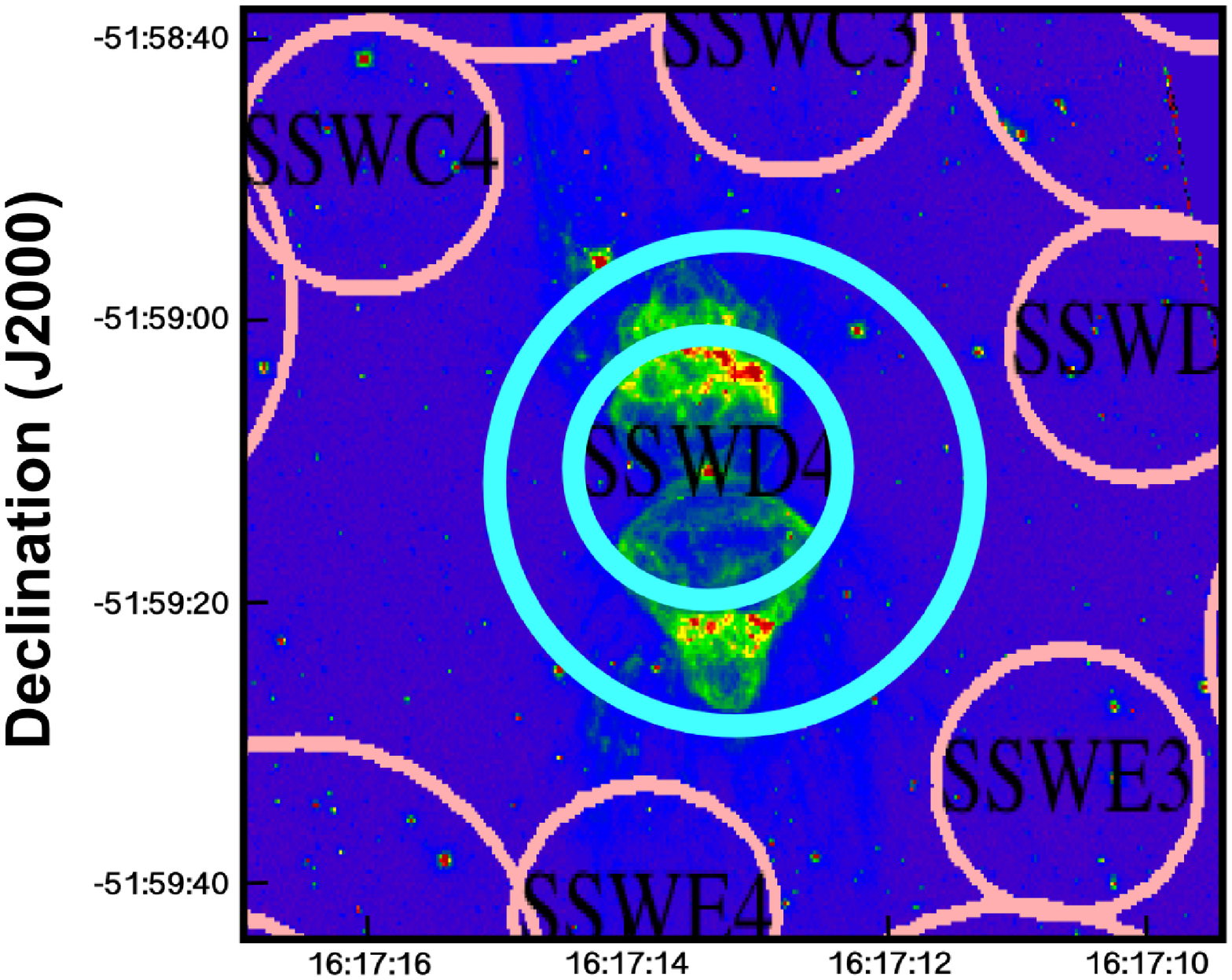}\\
   \includegraphics[width=8.2cm]{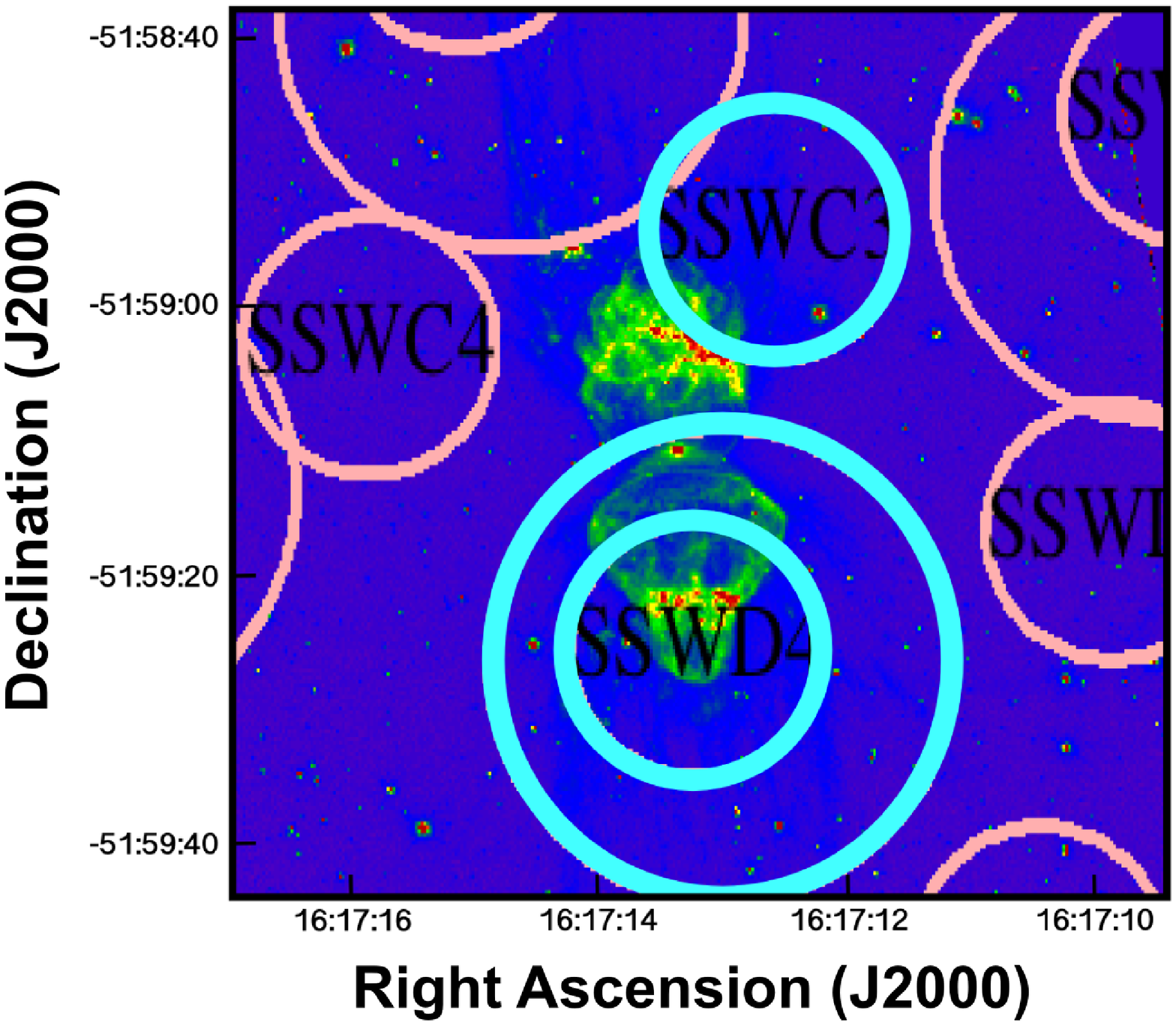}\\
   \caption{Mz~3 \textit{HerPlaNS} footprints for PACS (top panel) and SPIRE (centre pointing in the middle and southern lobe pointing in the bottom panel) observations plotted on a WFPC2/HST Mz~3 image using the H$\alpha$--[N~{\sc ii}] filter (F658N; GTO program PID 6856, P.I. J. Trauger). The SPIRE bolometers discussed in the text are highlighted in cyan; the others are represented in pink.}
   \label{footprints}
\end{figure}

The FIR to submm spectrum of Mz~3 presented in this paper was obtained by the \textit{Herschel Planetary Nebulae Survey} \citep[\textit{HerPlaNS}; ][]{2014A&A...565A..36U}. The survey acquired FIR/submm spectra and broadband images of eleven PNe with the PACS \citep[Photodetector Array Camera and Spectrometer; ][]{2010A&A...518L...2P} and SPIRE \citep[Spectral and Photometric Imaging Receiver; ][]{2010A&A...518L...3G} instruments on board the \textit{Herschel Space Observatory} \citep{2010A&A...518L...1P}. The resolving power of PACS and SPIRE depend on the wavelength. The resolving power ($\lambda / \Delta\lambda$) of PACS ranges from 1000 to 5500, while for SPIRE the resolving power ranges from 370 to 1288. PACS observations were taken for just one pointing toward the centre of the nebulae, while SPIRE observations were made for two pointings, one towards the centre and the other towards the southern lobe of Mz~3. The PACS spaxels and SPIRE bolometers footprints for each pointing are displayed in Fig. \ref{footprints}.

The data reduction procedure for the \textit{HerPlaNS} observations is described in \citet{2014A&A...565A..36U}. For PACS data, the reduction was performed with \textit{HIPE}\footnote{HIPE is a joint development by the Herschel Science Ground Segment Consortium, consisting of ESA, the NASA Herschel Science Center, and the HIFI, PACS and SPIRE consortia \citep{2010ASPC..434..139O}.} (version~11\footnote{Tests we performed showed that the differences in the fluxes found with the more recent version of \textit{HIPE} (version~15) are within the quoted uncertainties.}, calibration release version 44), using the background normalisation PACS spectroscopy pipeline script and following the procedure described in the \textit{PACS Data Reduction Guide: Spectroscopy}\footnote{\url{http://herschel.esac.esa.int/hcss-doc-9.0/load/pacs_spec/html/pacs_spec.html} (Version 1, Aug. 2012)}. For SPIRE, we used \textit{HIPE} (version~11, calibration tree version~11), following the standard HIPE-SPIRE spectroscopy data reduction pipeline for the single-pointing mode described in the \textit{SPIRE Data Reduction Guide}\footnote{\url{http://herschel.esac.esa.int/hcss-doc-9.0/load/spire_drg/html/spire_drg.html} (Version 2.1, Document Number: SPIRE-RAL-DOC 003248, 06 July 2012)}.

Line intensities were measured using the code {\sc HerFit}\footnote{The code, developed by I. Aleman, is available from the author upon request.}. The code fits the lines with Gaussian profiles. The continuum emission around the line is fitted with a polynomial curve with a user-defined degree. In our case, although a first or second degree polynomial was sufficient for several line fittings, assuming a third-degree polynomial showed the best results for all the lines when compared to {\sc Splat/Starlink} \citep{2014A&C.....7..108S} measurements. {\sc HerFit} is based on the MPFIT algorithm \citep{2009ASPC..411..251M, More1978}, which is a widely used algorithm that uses the Levenberg-Marquardt technique to solve the least-squares problem. {\sc HerFit} yields very similar results to already well established tools such as {\sc Splat/Starlink} and {\sc IRAF}/{\sc Splot}\footnote{IRAF is distributed by the National Optical Astronomy Observatories, which are operated by the Association of Universities for Research in Astronomy, Inc., under cooperative agreement with the National Science Foundation.}.

\section{Mz~3 PACS and SPIRE Spectra} \label{firspec}

\begin{figure*}
   \centering
   \includegraphics[width=15cm]{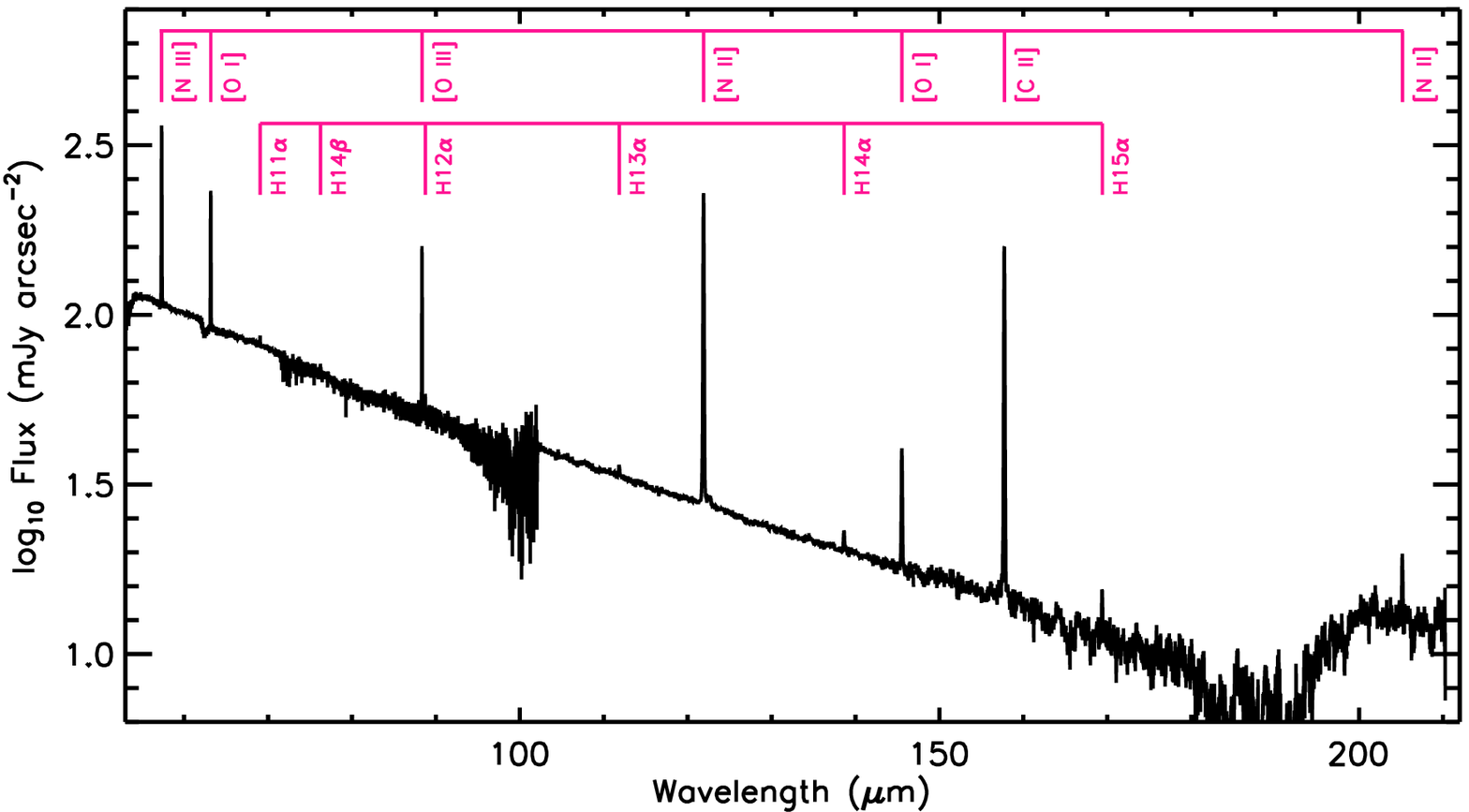}
   \includegraphics[width=15cm]{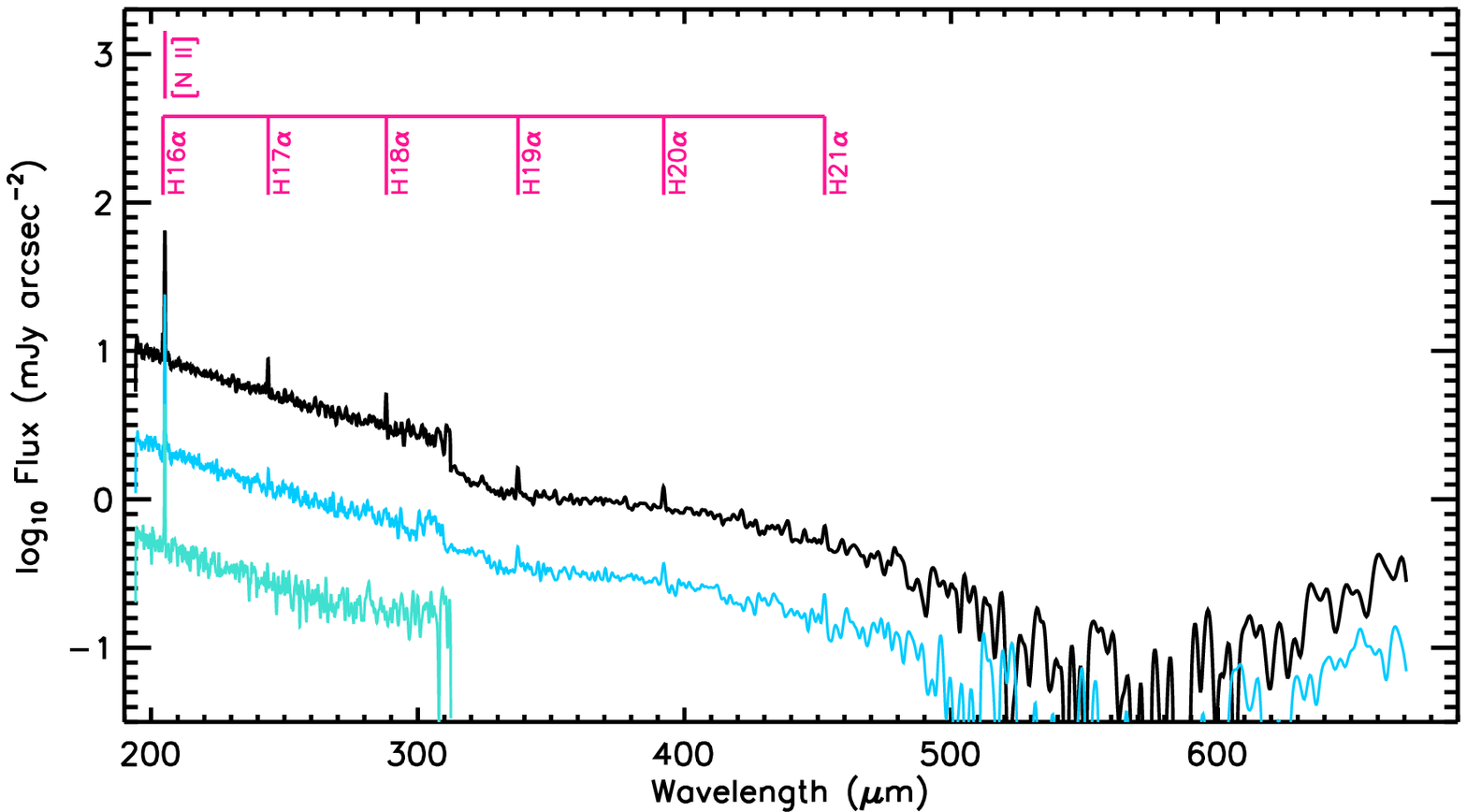}
   \caption{\textit{Top panel:} Integrated PACS spectrum of Mz~3. The spectrum was obtained by summing the flux of spectra extracted from all individual PACS spaxels. \textit{Bottom panel:} SPIRE spectra of Mz~3. From the top to bottom, the curves correspond, respectively, to the spectrum from the central bolometers for the central and southern lobe pointings (combined central SSW and SLW bolometers), and the bolometer SSWC3 for the southern lobe pointing. The wavelengths of the atomic forbidden lines, and H$n\alpha$ and H$n\beta$ lines detected are indicated.}
   \label{herschel_spec}
\end{figure*}

The full integrated PACS FIR spectrum of Mz~3 is shown in the top panel of Fig. \ref{herschel_spec}. The spectrum was obtained by integrating the flux of spectra extracted from all individual PACS spaxels. By simply summing the spaxel fluxes, we are not considering effects such as the point spread function (PSF) width exceeding the spaxel size. An uncertainty of 30 per cent has been added to the error in the flux measurements to account for this effect. 


The bottom panel of Figure \ref{herschel_spec} shows three SPIRE spectra obtained for Mz~3. From top to bottom, the figure shows the spectrum obtained with the central bolometer for the (i)~central and (ii)~southern lobe pointings, and (iii)~the bolometer SSWC3 for the southern lobe pointing (see Fig.\ref{footprints} for the footprints).


From Fig. \ref{herschel_spec}, it can be seen that the FIR spectrum of Mz~3 is dominated by atomic forbidden lines and a strong continuum due to thermal dust emission. Typical of PNe, the forbidden lines of [N~{\sc iii}] 57~$\mu$m, [O~{\sc i}] 63 and 145~$\mu$m, [O~{\sc iii}] 88~$\mu$m, [N~{\sc ii}] 122 and 205~$\mu$m, [C~{\sc ii}] 157~$\mu$m are present in the spectra. The [C{\sc i}] 370 and 609~$\mu$m lines were not detected.


\begin{table*}
\centering
\caption{Surface Brightness of the lines detected in the Mz~3 PACS spectrum}
\label{f_pacs}
\begin{tabular}{cccccccc}
\hline 
\multicolumn{2}{c}{Line}  & \multicolumn{2}{c}{All Spaxels} & \multicolumn{2}{c}{Three Spaxels} & \multicolumn{2}{c}{Central Spaxels}\\
  & $\lambda _{0}$ & $\lambda _{\rm obs}$ & $I$ & $\lambda _{\rm obs}$ & $I$ & $\lambda _{\rm obs}$ & $I$\\
\hline             
\multicolumn{8}{l}{Forbidden Lines}\\

[N~{\sc iii}] & 57.34  & 57.32  & 1543 $\pm$ 482 & 57.32  & 6205 $\pm$ 1992 & 57.32  & 4319 $\pm$ 1396 \\

[O~{\sc i}]   & 63.19  & 63.17  & 651  $\pm$ 205 & 63.17  & 2654 $\pm$ 835  & 63.17  & 3745 $\pm$ 1191 \\

[O~{\sc iii}] & 88.36  & 88.34  & 244  $\pm$ 81  & 88.35  & 860  $\pm$ 282  & 88.35  & 701  $\pm$ 246  \\

[N~{\sc ii}]  & 121.80 & 121.89 & 591  $\pm$ 186 & 121.89 & 1706 $\pm$ 550  & 121.89 & 1495 $\pm$ 483  \\

[O~{\sc i}]   & 145.54 & 145.52 & 47   $\pm$ 15  & 145.53 & 166  $\pm$ 54   & 145.52 & 211  $\pm$ 71   \\

[C~{\sc ii}]  & 157.68 & 157.73 & 257  $\pm$ 85  & 157.73 & 684  $\pm$ 228  & 157.73 & 830  $\pm$ 280  \\

\\
\multicolumn{8}{l}{Hydrogen Recombination Lines}\\

H11$\alpha$ & 69.07  & 69.06  & 14 $\pm$ 6  & 69.06  & 86 $\pm$ 31 & 69.06  & 166 $\pm$ 59 \\  
H14$\beta$  & 76.25  & 76.24  & 20 $\pm$ 10 & 76.25  & 60 $\pm$ 29 & 76.24  & [42  $\pm$ 26] \\
H12$\alpha$ & 88.75  & 88.74  & 7  $\pm$ 4  & 88.74  & 60 $\pm$ 27 & 88.74  & 143 $\pm$ 59 \\  
H13$\alpha$ & 111.86 & 111.84 & 8  $\pm$ 3  & 111.83 & 52 $\pm$ 19 & 111.83 & 116 $\pm$ 41 \\  
H14$\alpha$ & 138.65 & 138.63 & 8  $\pm$ 3  & 138.63 & 46 $\pm$ 16 & 138.63 & 94  $\pm$ 32 \\  
H15$\alpha$ & 169.41 & 169.39 & 5  $\pm$ 3  & 169.39 & 30 $\pm$ 13 & 169.39 & 60  $\pm$ 26 \\

\hline      

\multicolumn{8}{l}{Notes: Rest and observed wavelengths ($\lambda_{0}$ and $\lambda _{\rm obs}$) given in $\mu$m and surface brightness}\\

\multicolumn{8}{l}{ ($I$) in 10$^{-17}$~erg~cm$^{-2}$~s$^{-1}$~arcsec$^{-2}$. Rest wavelengths ($\lambda_{0}$) from \citet{NIST_ASD}.}\\

\multicolumn{8}{l}{The H14$\beta$ central spaxel surface brightness is indicated \textbf{in} brackets to indicate a}\\

\multicolumn{8}{l}{low signal-to-noise ratio (S/N = 2.6).}\\

\end{tabular}
\end{table*}

\begin{table*}
\centering
   \caption{Mz~3 lines detected with SPIRE (central bolometer flux only)}
   \label{f_spire}
\begin{tabular}{ccccccc}
\hline
\multicolumn{2}{c}{Line} & \multicolumn{2}{c}{Centre} & \multicolumn{2}{c}{Southern Lobe} & Beam\\
 Carrier & $\lambda _{0}$ & $\lambda _{\rm obs}$ & $I$ & $\lambda _{\rm obs}$ & $I$ & FWHM \\    
\hline             
\multicolumn{4}{l}{Forbidden Line}\\
 
[N~{\sc ii}]      & 205.30 & 205.17 & 130  $\pm$ 2    & 205.16 & 128  $\pm$ 3    & 17 \\  
\\
\multicolumn{4}{l}{Hydrogen Recombination Lines}\\

H16$\alpha$ & 204.41 & 204.38 & 8.9  $\pm$ 0.5  & 204.43 & 1.2  $\pm$ 0.2  & 17 \\
H17$\alpha$ & 243.93 & 243.86 & 8.3  $\pm$ 0.4  & 244.02 & 3.1  $\pm$ 0.3  & 17 \\
H18$\alpha$ & 288.23 & 288.14 & 4.6  $\pm$ 0.2  & 288.03 & 2.7  $\pm$ 0.3  & 19 \\
H19$\alpha$ & 337.59 & 337.57 & 1.16 $\pm$ 0.06 & 337.56 & 0.99 $\pm$ 0.05 & 36 \\
H20$\alpha$ & 392.28 & 392.16 & 0.73 $\pm$ 0.02 & 392.25 & 0.72 $\pm$ 0.05 & 34 \\
H21$\alpha$ & 452.58 & 452.65 & 0.33 $\pm$ 0.02 & 452.47 & 0.44 $\pm$ 0.06 & 30 \\
\hline

\multicolumn{7}{l}{Notes: Rest and observed wavelengths ($\lambda_{0}$ and $\lambda _{\rm obs}$) given in $\mu$m, surface}\\

\multicolumn{7}{l}{brightness in 10$^{-17}$~erg~cm$^{-2}$~s$^{-1}$~arcsec$^{-2}$, and beam full width at half}\\

\multicolumn{7}{l}{maximum (FWHM) in arcsec \citep{2010A&A...518L...4S}. Rest wavelengths}\\

\multicolumn{7}{l}{from \citet{NIST_ASD}.}\\

\end{tabular}
\end{table*}


The surface brightness of individual detected lines are presented in Tables \ref{f_pacs} and \ref{f_spire}. The first table lists the line surface brightnesses obtained with PACS. Measurements are provided for three distinct regions: (i) all the PACS spaxels, (ii) the combined spaxels [1,2], [2,2] and [3,2], and (iii) the single central spaxel [2,2]. The area of each spaxel is 9.4$\arcsec$~$\times$~9.4$\arcsec$. The integrated surface brightness measurements from all PACS spaxels are similar to the values obtained by \citet{2005A&A...444..861P} and \citet{2001MNRAS.323..343L} from the Mz~3 \textit{ISO} spectrum. The \textit{ISO} and \textit{Herschel}/PACS FOV and pointing characteristics are not equal, but both \textbf{the} \textit{ISO} observations and our PACS spectra cover \textbf{the} lobes and the central region, where most of the forbidden line emission is produced. 

The [N~{\sc ii}] 205~$\mu$m line is present in both PACS and SPIRE spectra. The quality of the SPIRE spectrum around 205 $\mu$m is much better than the corresponding spectral region of the PACS spectrum. The region longwards of 190 microns is not well calibrated in the PACS data produced with HIPE version 11. Therefore we recommend the use of the [N~{\sc ii}] 205~$\mu$m line surface brightness measured from the SPIRE spectrum.

Table \ref{f_spire} presents the measurements from the SPIRE central bolometer for each pointing, i.e. centre and southern lobe. The [N~{\sc ii}] 205 $\mu$m line surface brightnesses measured in each SPIRE pointings are very similar. 


For the forbidden lines, between 32 and 49 per cent of the total flux collected by PACS comes from the three-spaxel region mentioned above, while between 8 and 23 per cent comes from the central spaxel region. This indicates that the lobes produce a large fraction of the total Mz~3 forbidden line emission. \citet{2003MNRAS.342..383S} showed that the emission of some lines, namely [N~{\sc ii}]~6583~\AA, [O~{\sc ii}]~3727~\AA, and [S~{\sc ii}]~6717+6731~\AA, is almost absent in the nucleus. Such emission comes mostly from the lobes, while other lines, such as [O~{\sc iii}]~5007~\AA~ and [S~{\sc iii}]~9068~\AA, have important contributions from the two lobes and the core. H\,$\alpha$ emission is also found to originate from the core and the two lobes \citep[see Fig. \ref{footprints} and][]{2003MNRAS.342..383S}.


The lines above show that the gas in the lobes is mostly ionized and has an ionization structure that radially extends from the mostly ionized plasma (e.g. [O~{\sc iii}]) to a surrounding low-ionization gas (e.g. [O~{\sc i}]). \citet{2003MNRAS.342..383S} studied the atomic emission from the lobes and determined their average physical conditions. From the line ratio empirical analysis, he inferred an electron density of $n_{\rm e} \sim$ 4500~cm$^{-3}$ and electronic temperatures in the interval $T_{\rm e} \sim$~7000--15000~K (depending on the ion used). Similar values were found by \citet{2005A&A...444..861P}. Using photodissociation region (PDR) line diagnostics diagrams \citep[The PDR Toolbox\footnote{\url{http://dustem.astro.umd.edu/pdrt/}};][]{2006ApJ...644..283K,2008ASPC..394..654P}, the forbidden line ratios derived from the measurements in Table \ref{f_pacs} (PACS range lines only) provide the following gas densities ($n_{\rm H}$) and incident fluxes \citep[$G_0$, the incident flux in Habing units, i.e., relative to the average interstellar medium flux of 1.6~$\times$~10$^{-3}$~ergs~cm$^{-2}$~s$^{-1}$; ][]{2008ASPC..394..654P}: $n_{\rm H}$~=~1800, 5600, 5600~cm$^{-3}$ and $G_0$~=~3200, 1000, 1800 for, respectively, all spaxels, the three central spaxels, and the central spaxel measurements. A detailed pan-chromatic photoionization model \citep[e.g.][]{2017ApJS..231...22O} of Mz~3 will be presented in a upcoming \textit{HerPlaNS} paper.

No molecular line emission was detected in our data set. The absence of CO emission in our spectra, even from the core spaxel, where a dense disc/torus is present, is noteworthy. Weak emission of CO $J =$~2--1 was detected by \citet{1991A&A...242..247B}, indicating a total molecular mass of 1.7~$\times$~10$^{-3}$ $M_{\sun}$, which corresponds to a fraction of 1.3~$\times$~10$^{-2}$ of the ionized mass, according to \citet{1996A&A...315..284H}. Although the detection of CO has been reported, no H$_2$ rovibrational lines are present in the near-IR spectrum obtained by \citet{2003MNRAS.342..383S} and no rotational H$_2$ lines are present in the ISO spectra published by \citet{2005A&A...444..861P}. 


The most striking feature in the Mz~3 \textit{Herschel} spectrum is, however, the presence of HRLs, which we discuss in the next section.

\section{H~{\sc i} Recombination Lines in Mz~3} \label{recomb}


We detected hydrogen recombination $\alpha$ lines from H11$\alpha$ to H21$\alpha$ in the \textit{Herschel} PACS and SPIRE spectra of Mz~3 (Fig. \ref{herschel_spec}). Only one $\beta$ HRL was detected in the range: H14$\beta$. 


To the best of our knowledge, no detection of HRLs in radio frequencies has been reported for Mz~3 thus far. In previous studies of this nebula with the \textit{ISO} mid-infrared (MIR) spectrum, the H4$\alpha$ ($n = 5\rightarrow 4$ 4.05$\mu$m) and H5$\alpha$ ($n = 6\rightarrow 5$ 7.46$\mu$m) lines were detected, but no lines from higher $n$ levels were seen \citep{2003ApJS..147..379S, 2005A&A...444..861P}. Many lines from the hydrogen series have been detected in the optical range \citep{2002MNRAS.337..499Z,2003MNRAS.342..383S}.


Surface brightnesses for the HRLs observed with \textit{Herschel} are listed in Tables \ref{f_pacs} and \ref{f_spire}. As described in the previous section, the first table shows the line surface brightnesses obtained with PACS, while the second presents the measurements with the SPIRE central bolometer for the two available pointings (centre and southern lobe). H$16\alpha$ at 204 $\mu$m was only detected with SPIRE. 

\begin{figure*}
   \centering
   \includegraphics[width=8.6cm]{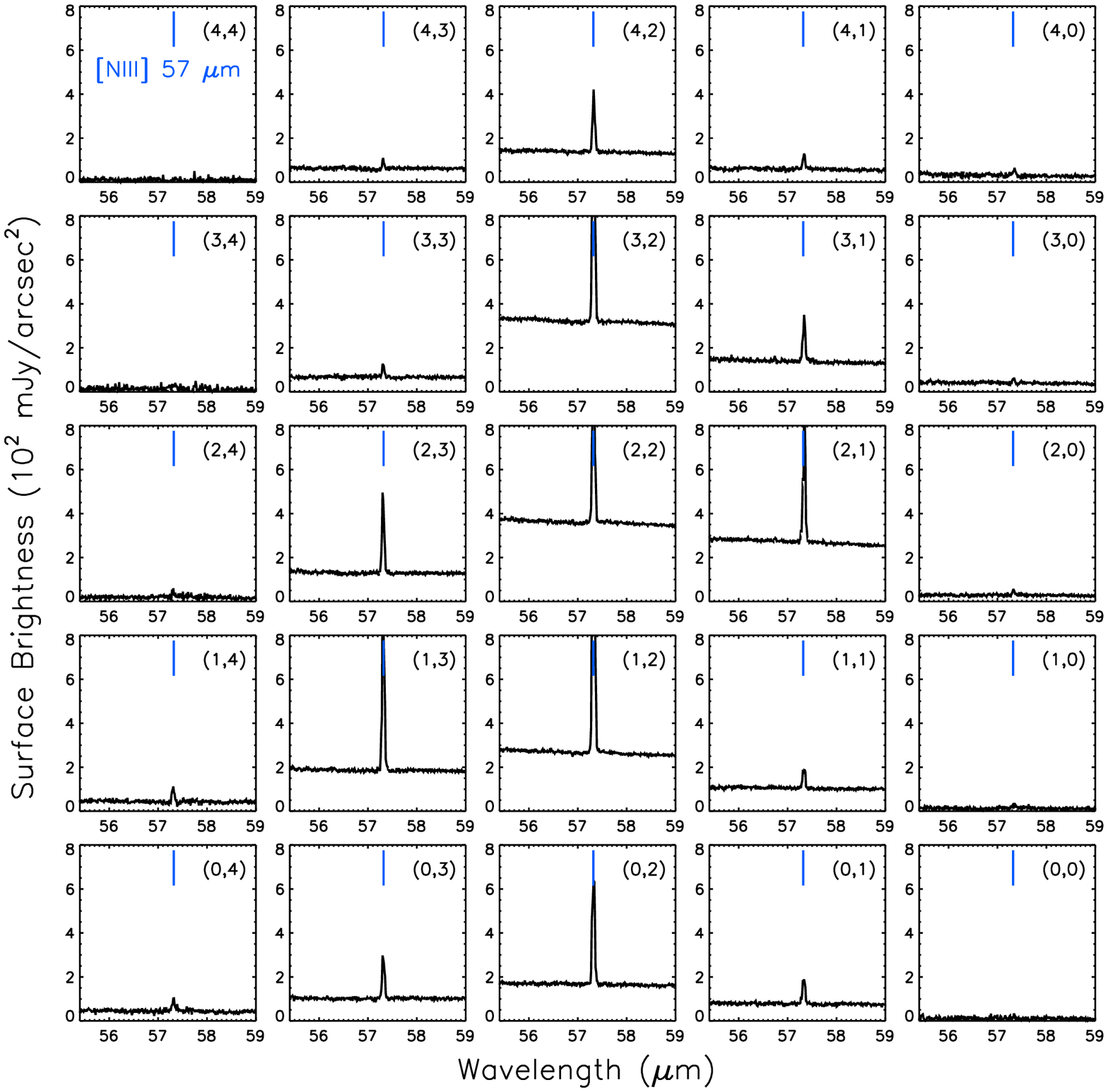}
   \includegraphics[width=8.6cm]{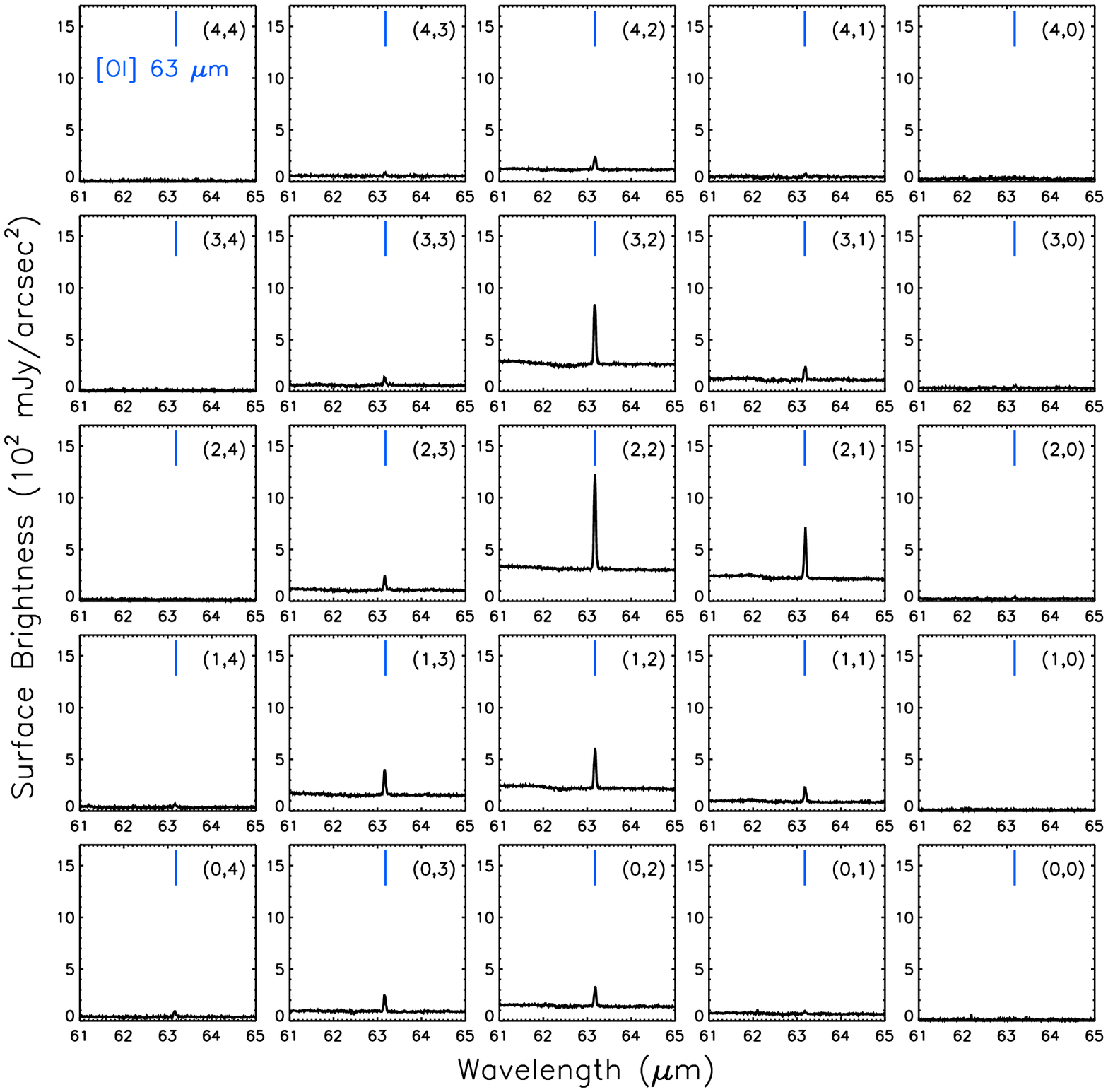}\\
   \includegraphics[width=8.6cm]{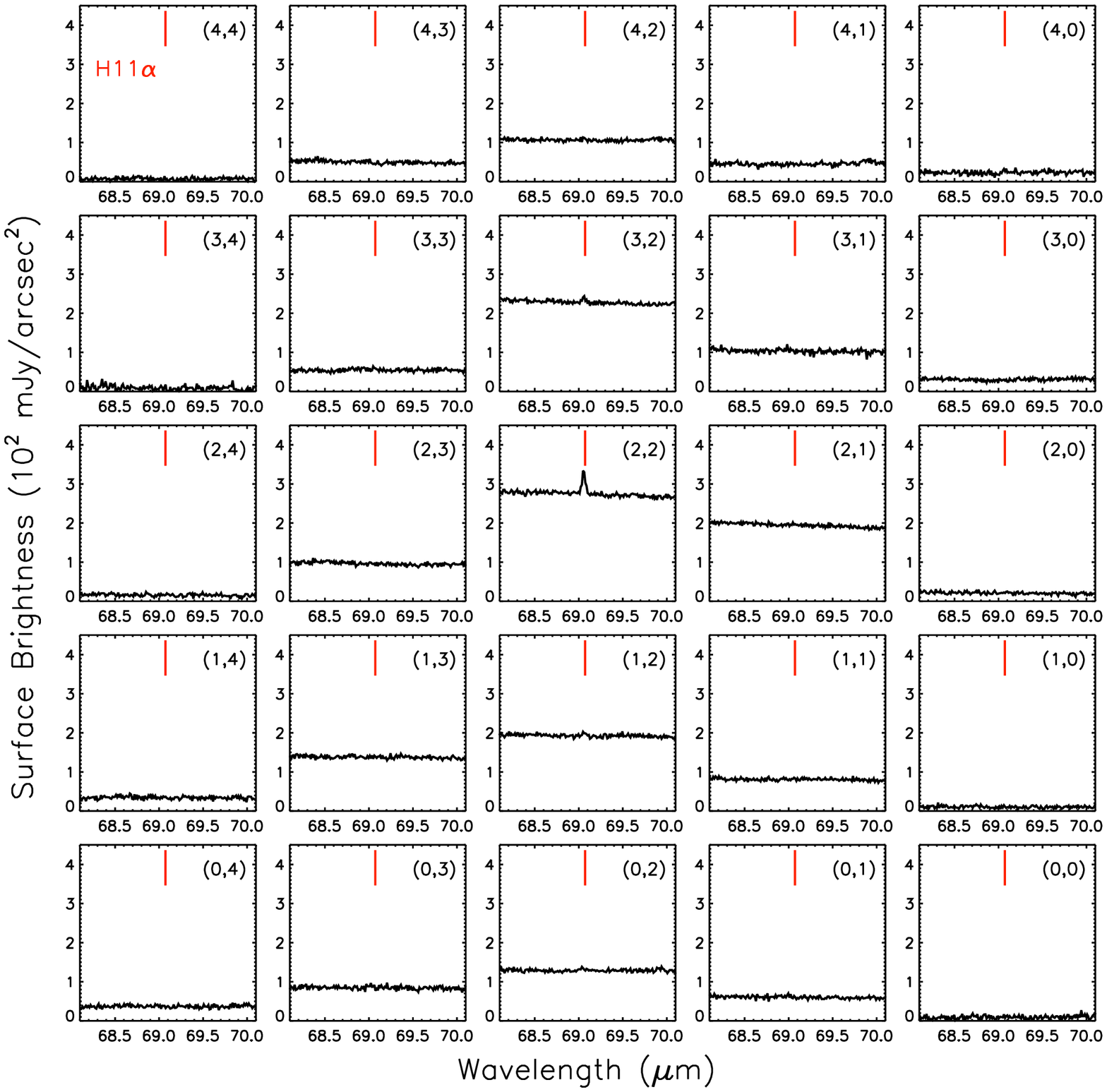}
   \includegraphics[width=8.6cm]{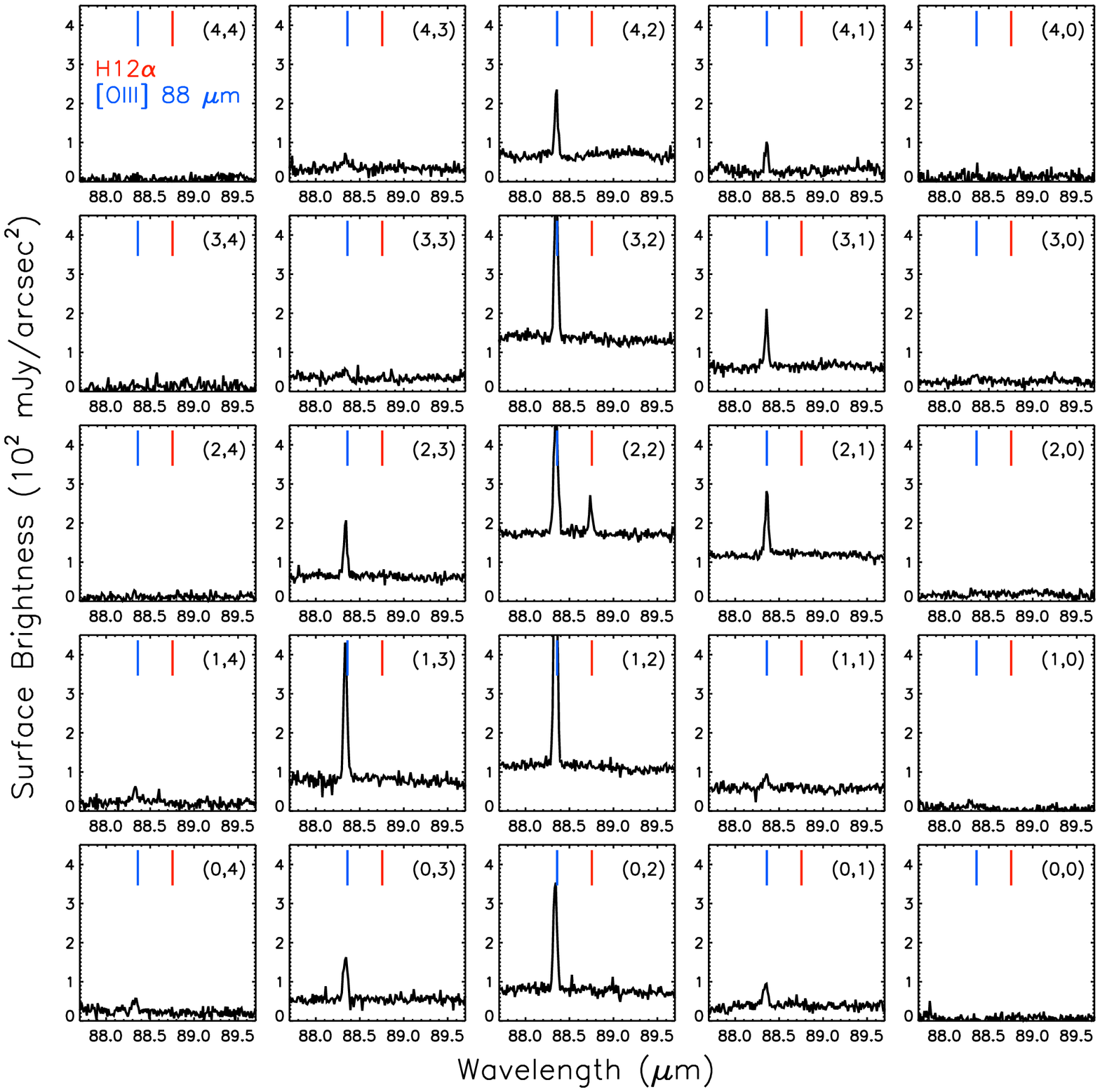}
   \caption{Spatial distribution of lines detected in the \textit{Herschel}/PACS FIR spectrum of Mz~3. The coordinates of the corresponding spaxel are indicated in the top right corner of each panel. The coordinates correspond to those indicated in the PACS footprint displayed in Fig. \ref{footprints}.}
   \label{spaxels1}
\end{figure*}
\begin{figure*}
   \centering
   \includegraphics[width=8.6cm]{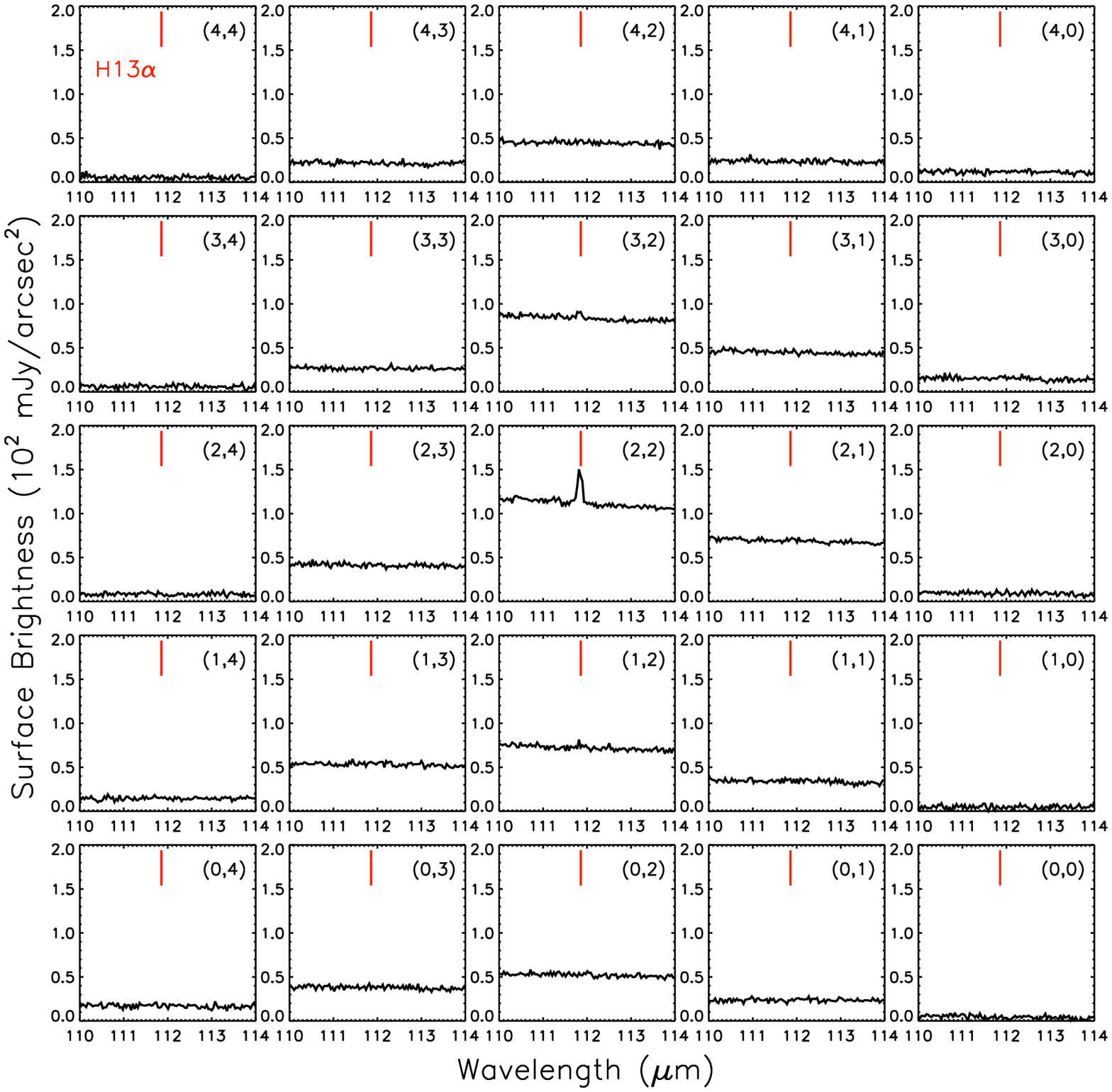}
   \includegraphics[width=8.6cm]{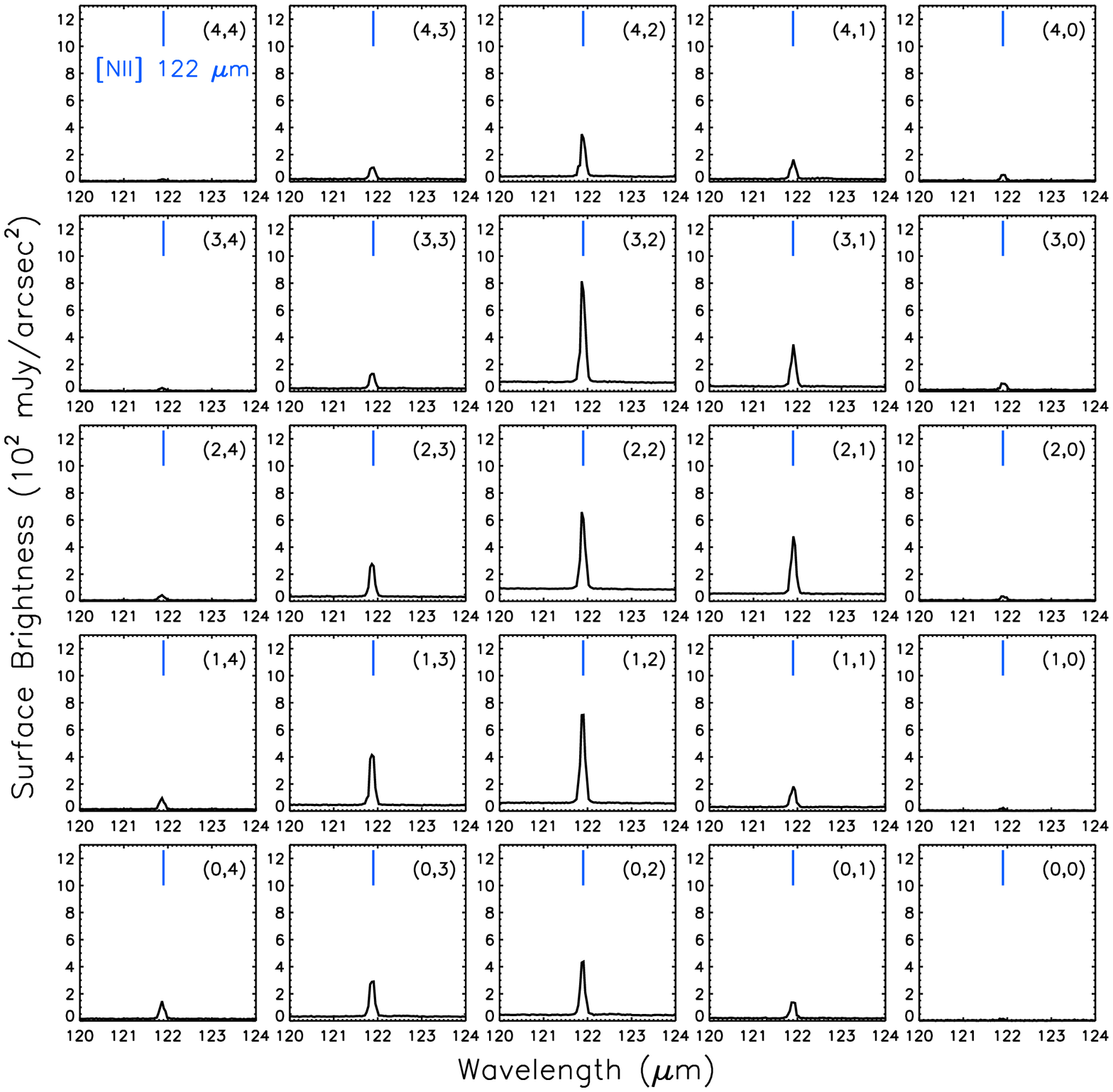}\\
   \includegraphics[width=8.6cm]{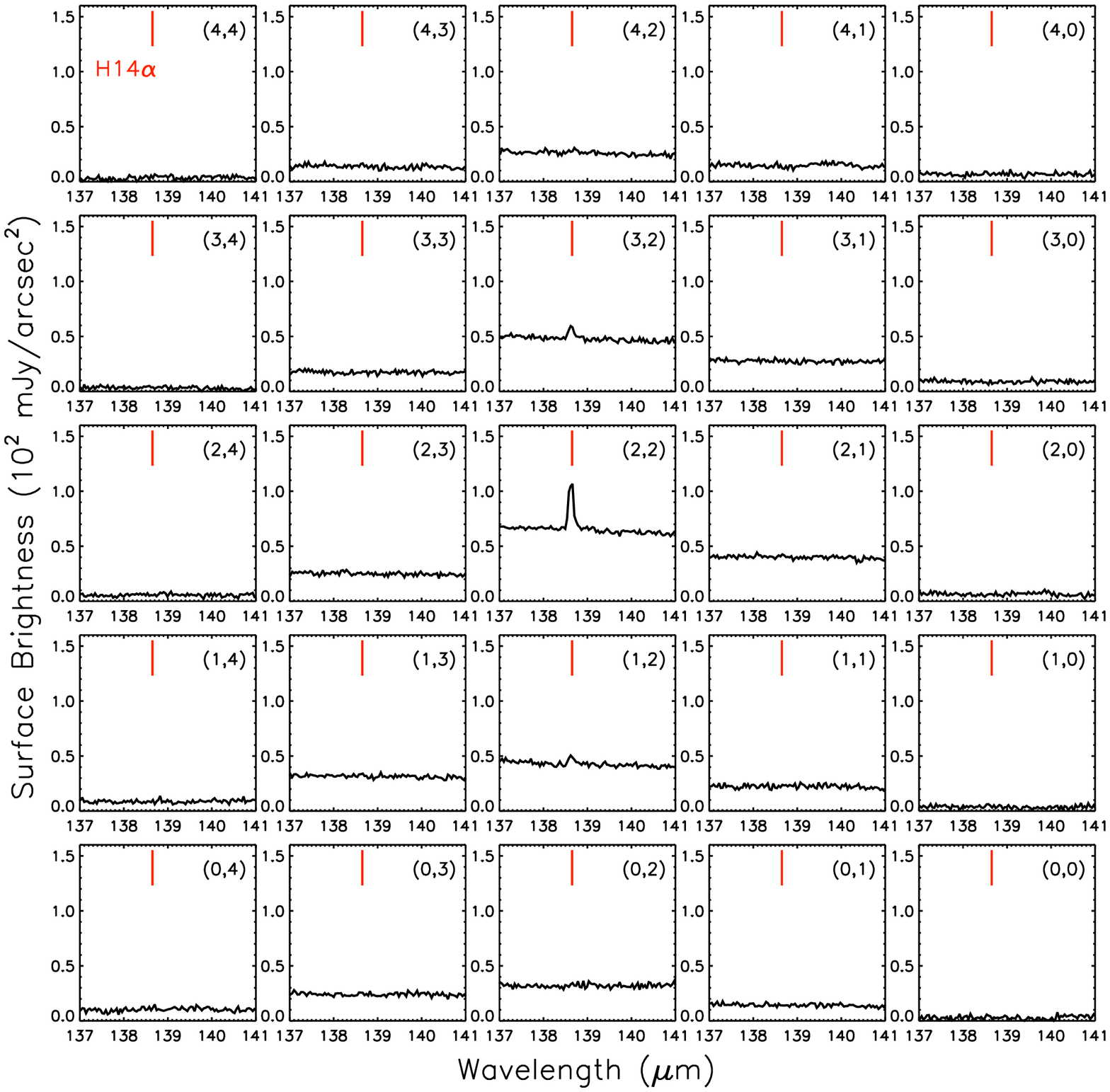}
   \includegraphics[width=8.6cm]{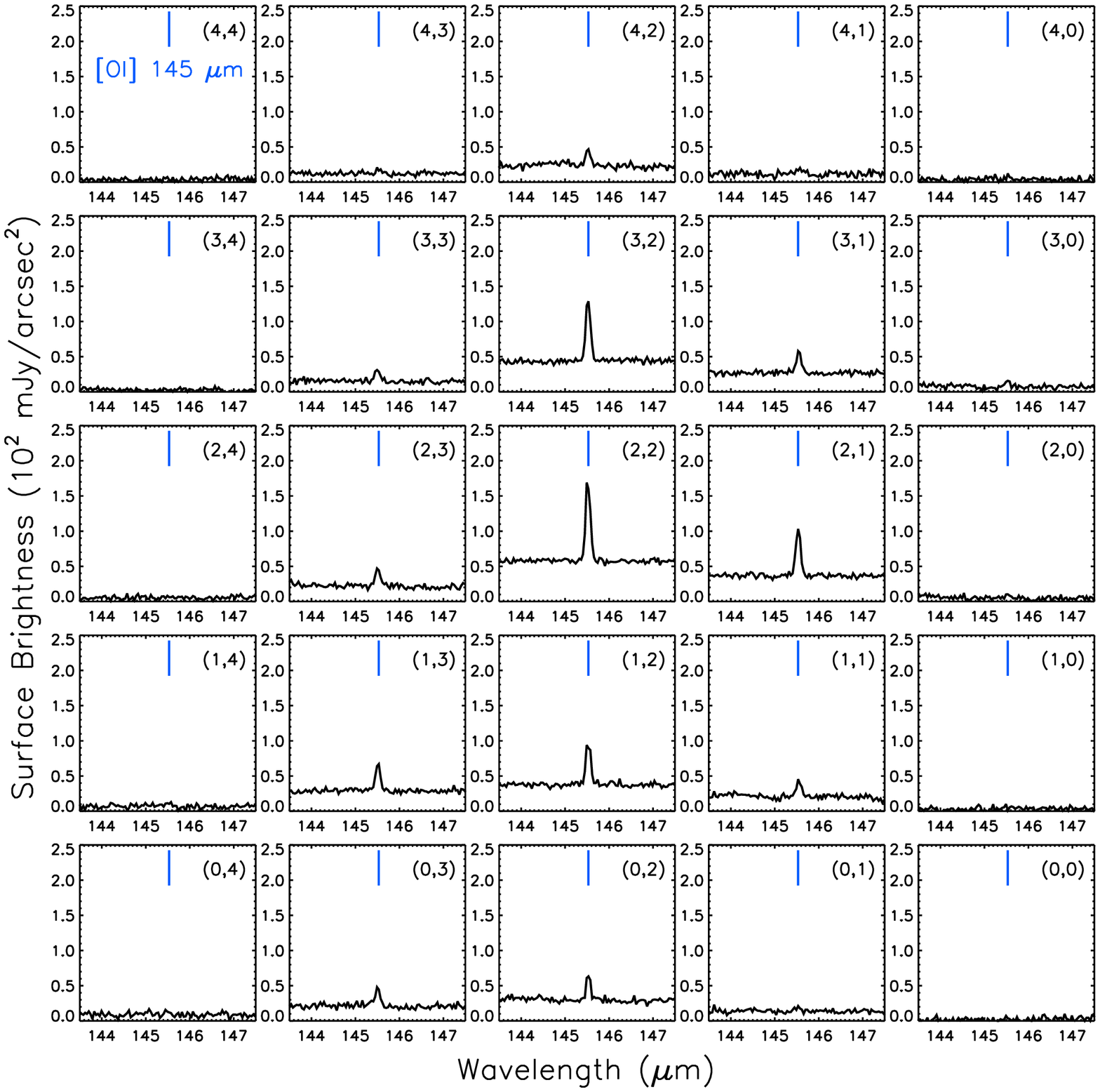}
   \caption{Same as Fig. \ref{spaxels1}, for different lines.}
   \label{spaxels2}
\end{figure*}
\begin{figure*}
   \centering
   \includegraphics[width=8.6cm]{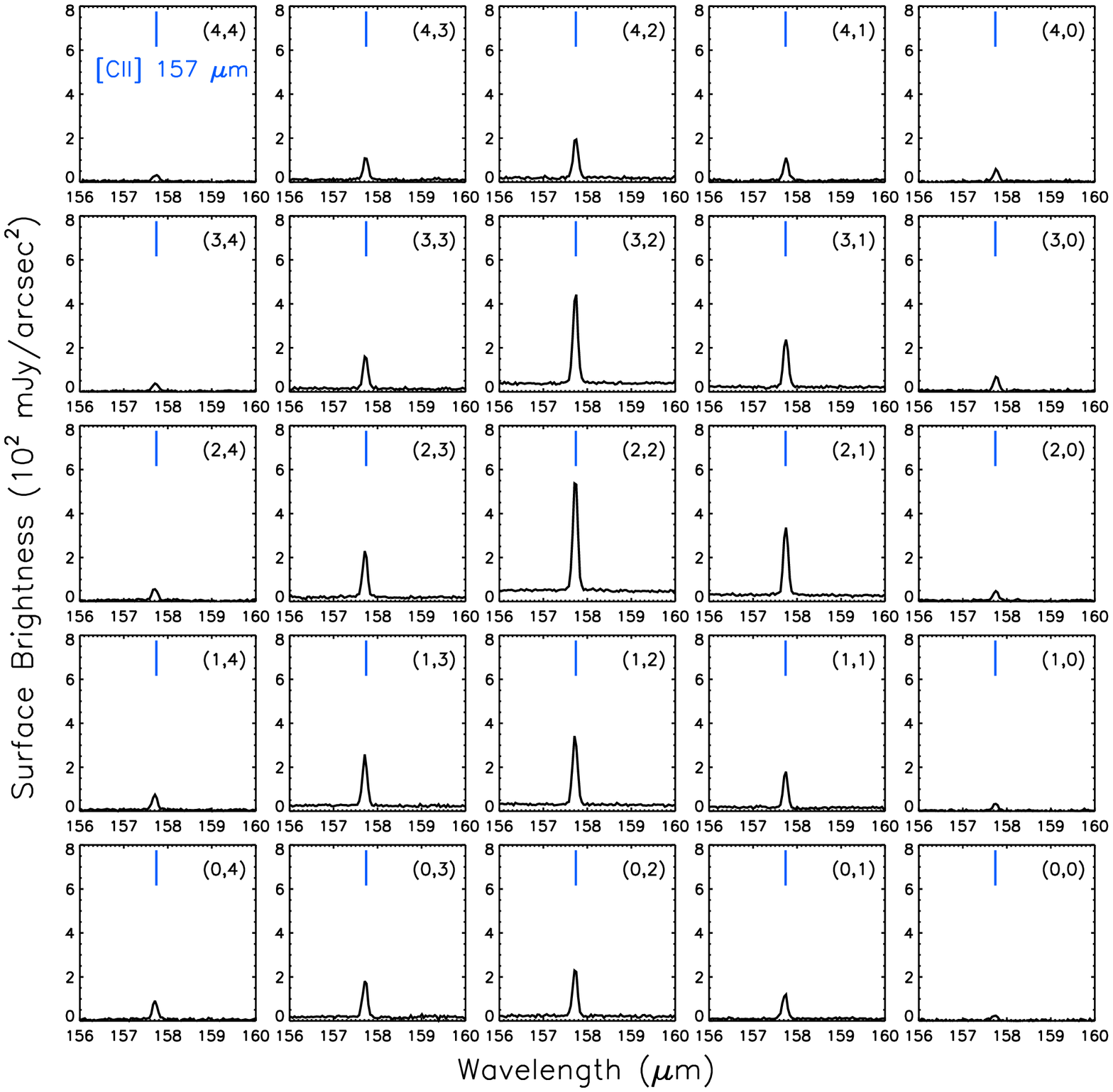}
   \includegraphics[width=8.6cm]{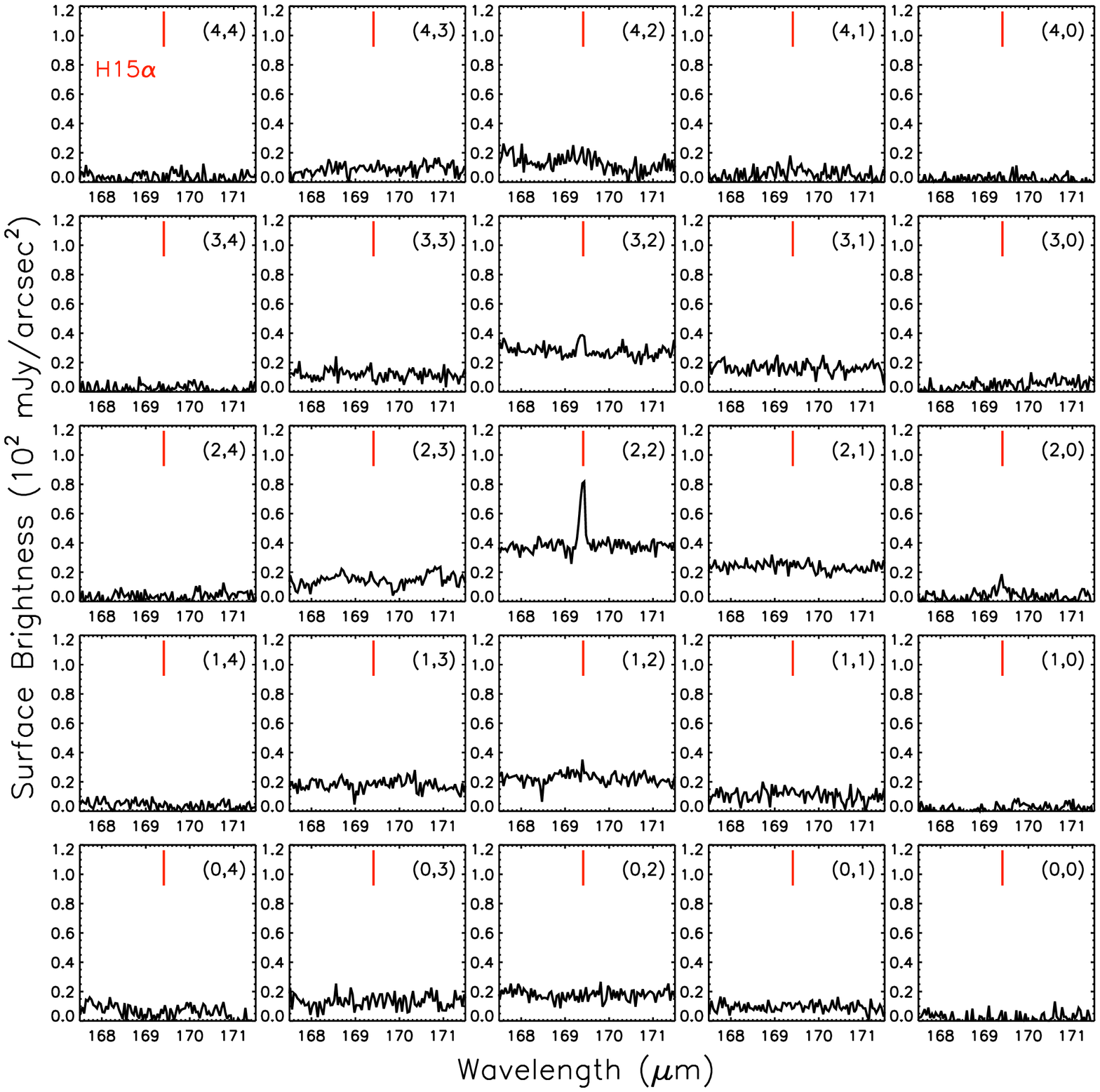}\\
   \includegraphics[width=8.6cm]{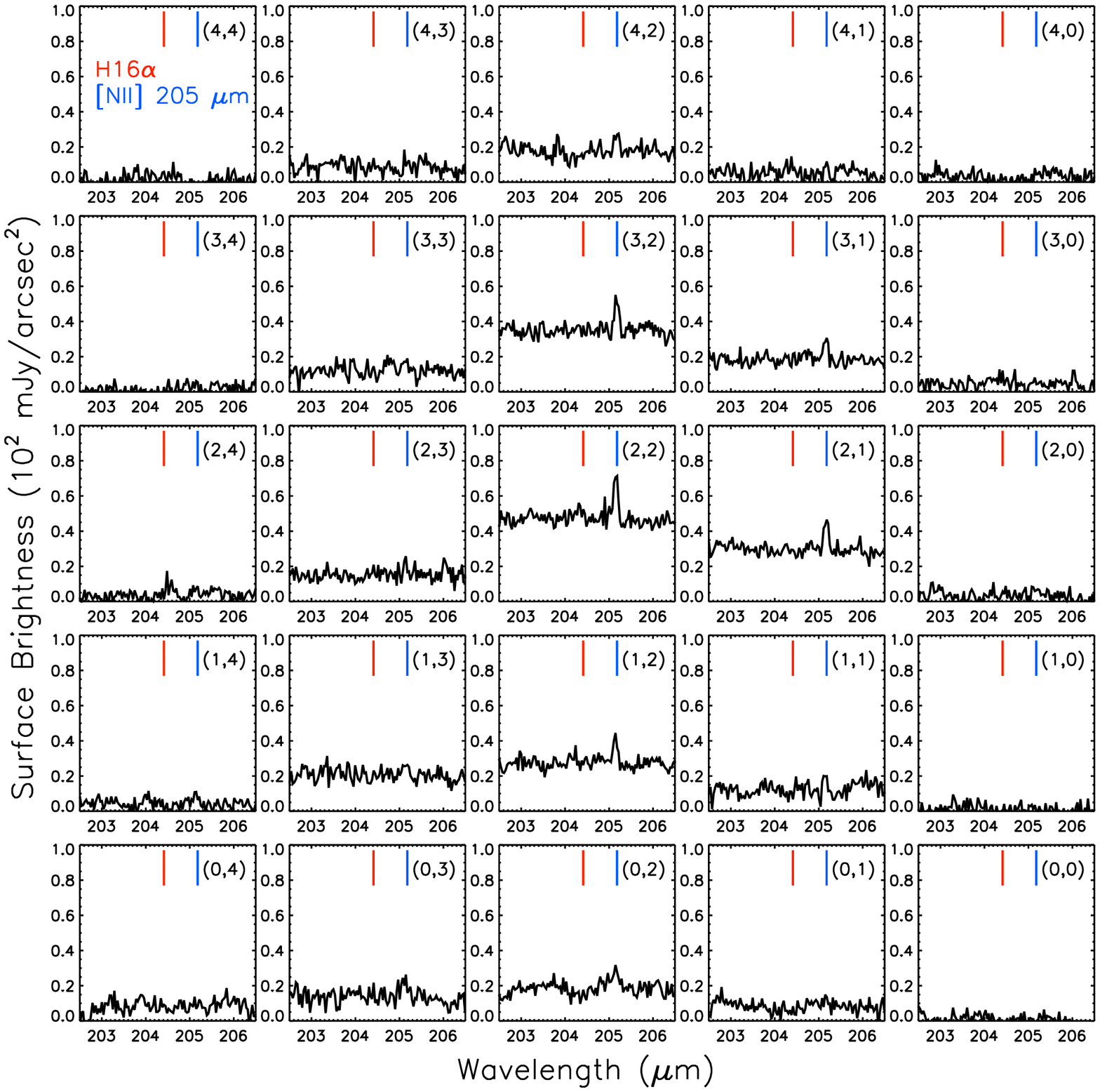}
   \caption{Same as Fig. \ref{spaxels1}, for different lines.}
   \label{spaxels3}
\end{figure*}

Figures \ref{spaxels1} to \ref{spaxels3} show the PACS spectra for each spaxel zoomed in around the lines detected in the nebula. The H$n\alpha$ recombination lines detected with \textit{Herschel} are mostly produced in the central region of Mz~3, while the C, N, and O forbidden lines, a significant fraction is produced in the lobes. The central spaxel, with some contribution from the neighbour spaxels [1,2] and [3,2] (each covering part of one of the Mz~3 lobes; see Fig. \ref{footprints}), dominates the emission of the HRLs. Intense emission of forbidden lines present typically in the ionized region (e.g. [O~{\sc iii}] 88~$\mu$m and [N~{\sc ii}] 122~$\mu$m lines) is much more extended than the emission of the FIR H$n\alpha$ lines in Mz~3. From Table \ref{f_pacs}, we see that the HRLs surface brightnesses of the central spaxel tend to be higher than those of the spectra integrated over the central three spaxels and over all the spaxels, indicating that the emission is highest close to the centre. For the forbidden lines, the central spaxel surface brightness is comparable (within uncertainties) to the value obtained from the integrated central three spaxels, but much smaller than when all the spaxels are integrated. 

The H14$\beta$ line at 76 $\mu$m is very faint and could not be detected above the three-sigma limit in individual spaxels. The line is, however, detected in the integrated spectra (all and three central spaxels).



It is interesting to compare the relative intensity of the of the PACS FIR HRLs at different positions in the nebula with those of the optical HRLs. \citet{2002MNRAS.337..499Z} measured the H\,$\beta$ emission in a slit positioned along the polar symmetry axis of the nebula, showing that the profile has three peaks, one in the core and one in each lobe. The core emission corresponds to $\sim$45 per cent of the total H\,$\beta$ emission. In our measurements of the H$15\alpha$ line at 139 $\mu$m, for example, the central spaxel emission corresponds to 47 per cent of the total PACS emission for this line. For other FIR H~{\sc i} lines, the central spaxel emission can account for 45 to 84 per cent of the total PACS emission. If we consider not only the central spaxel, but also spaxels [1,2] and [3,2], the emission corresponds to 70 per cent of the total PACS emission of H$15\alpha$ and 68 to 100 per cent for the other H$n\alpha$ lines. As a comparison, the fraction of the forbidden line emission from the central spaxel mentioned in the previous section ranges from 10 to 23 per cent and for the three central spaxels ranges from 32 to 49 per cent of the total PACS flux. For H14$\alpha$ line the fractions are 8 and 36 per cent for the central and three central spaxels, respectively.


From all the SPIRE bolometers, for both pointings, only the central bolometer (see footprints in Fig \ref{footprints}) shows HRL emission. For the centre pointing, the central bolometer covers the core and and part of the bipolar lobes depending on the wavelength, since the SPIRE bolometer beam size depends on the wavelength. In Fig \ref{footprints}, the size of the two circles representing the central bolometers indicates approximately the size of the beam for the smallest and largest possible SPIRE beams.

\begin{figure}
   \centering
   \includegraphics[width=7.8cm]{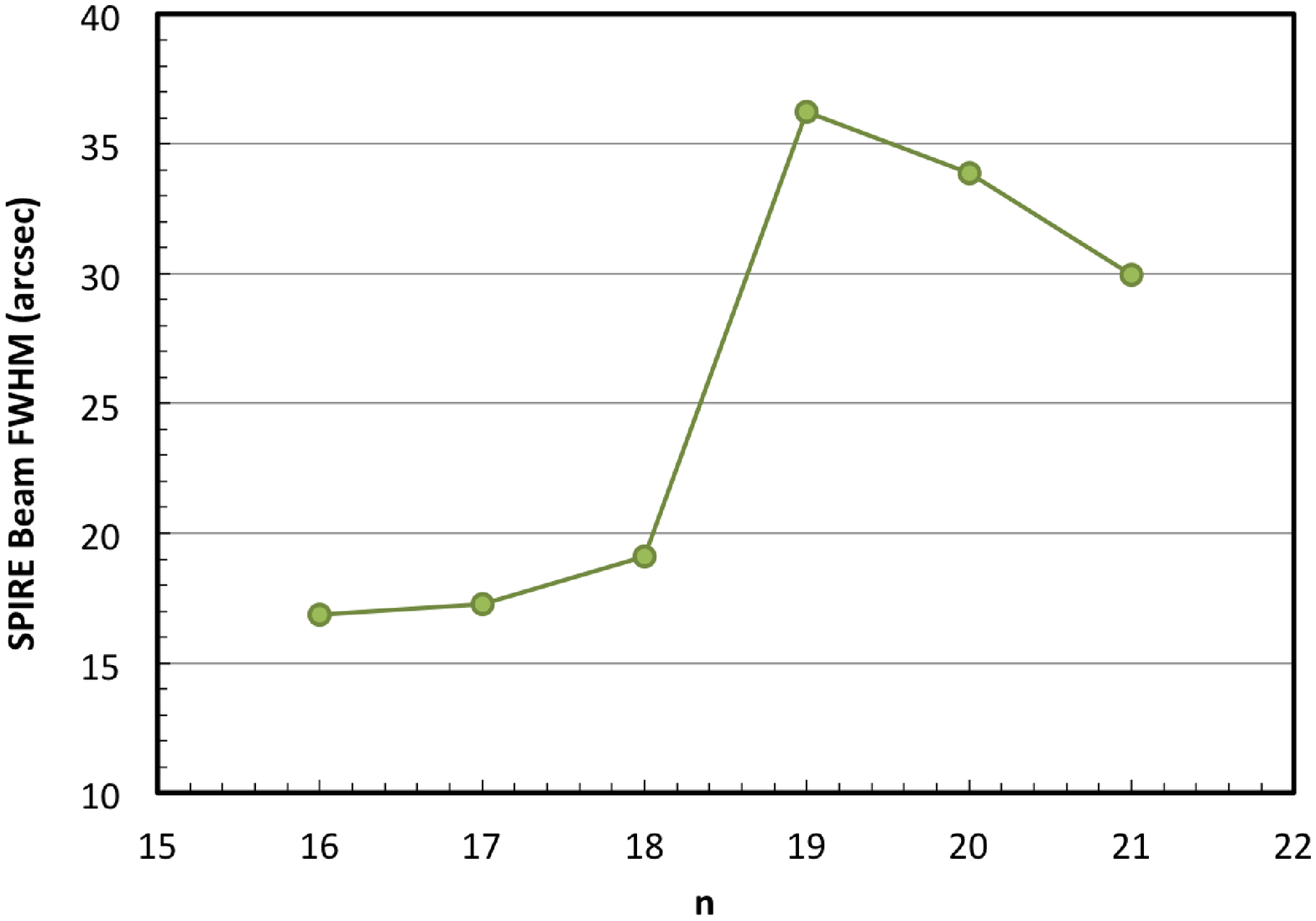}
   \caption{\textit{Herschel}/SPIRE beam size (full width at half maximum) for the wavelengths of each H$n\alpha$ line observed in Mz~3.}
   \label{spire_lines_beam}
\end{figure}

For the lobe pointing, the central bolometer observation covers part of the southern lobe and may include part of the central core depending on the wavelength observed. Figure \ref{spire_lines_beam} shows the beam sizes for each H$n\alpha$ line observed in our spectra \citep{2013ApOpt..52.3864M}. For shorter wavelengths, the central bolometer spectra should be dominated by the southern lobe emission.

In Fig. \ref{herschel_spec}, we also show the SSWC3 bolometer spectrum for the lobe pointing, which only covers short wavelengths. The bolometer covers the northern part of the northern lobe and does not show any evidence of HRL emission.

\section{Evidence of Laser Effect} \label{discuss}


\paragraph*{The detection of FIR/submm HRLs is unusual.}
The detection of HRLs in the FIR/submm spectrum of Mz~3 was not initially expected as in the range of physical conditions typical for PNe such lines should be much fainter than the atomic forbidden lines. Hydrogen recombination lines were not detected in any of the other 10 PNe observed in \textit{HerPlaNS} (NGC~40, NGC~2392, NGC~3242, NGC~6445, NGC~6543, NGC~6720, NGC~6781, NGC~6826, NGC~7009, and NGC~7026). A search in the literature revealed that FIR/submm HRLs have been observed in only a few PNe. Such lines have been reported in the young PN NGC~7027 by \citet{1996A&A...315L.257L} and \citet{2010A&A...518L.144W} and in the pre-PNe AFGL~2688 and AFGL~618 by \citet{2010A&A...518L.144W}. The identification of H12$\alpha$ and H13$\alpha$ made by \citet{1996A&A...315L.257L} in the ISO FIR spectrum of the young PN NGC~7027 is questionable according to the authors themselves. However, \citet{2010A&A...518L.144W} reported the detection of a few HRLs in the SPIRE spectra of NGC~7027.

\begin{figure}
   \centering
   \includegraphics[width=8.1cm]{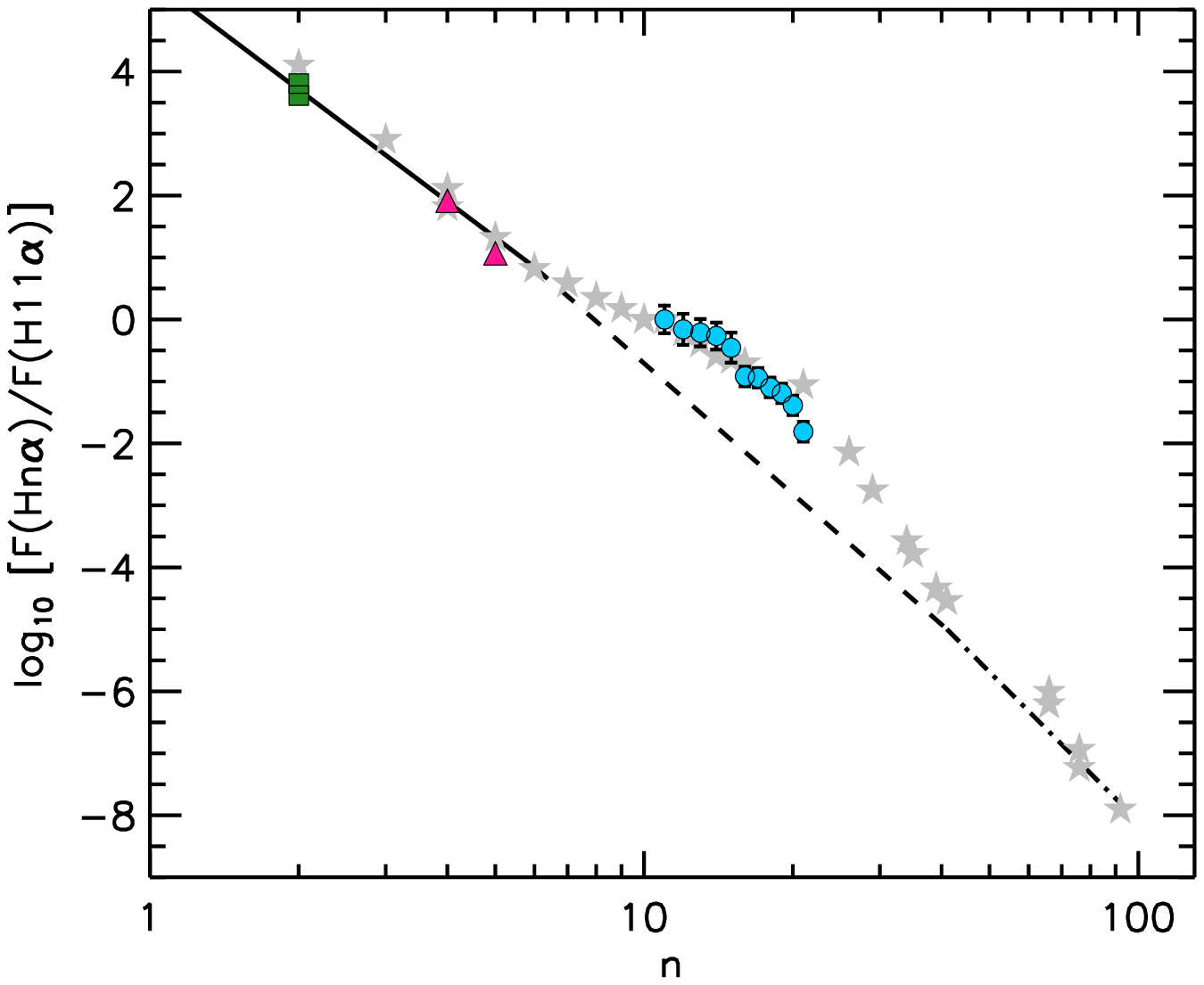}\\
   \includegraphics[width=8.1cm]{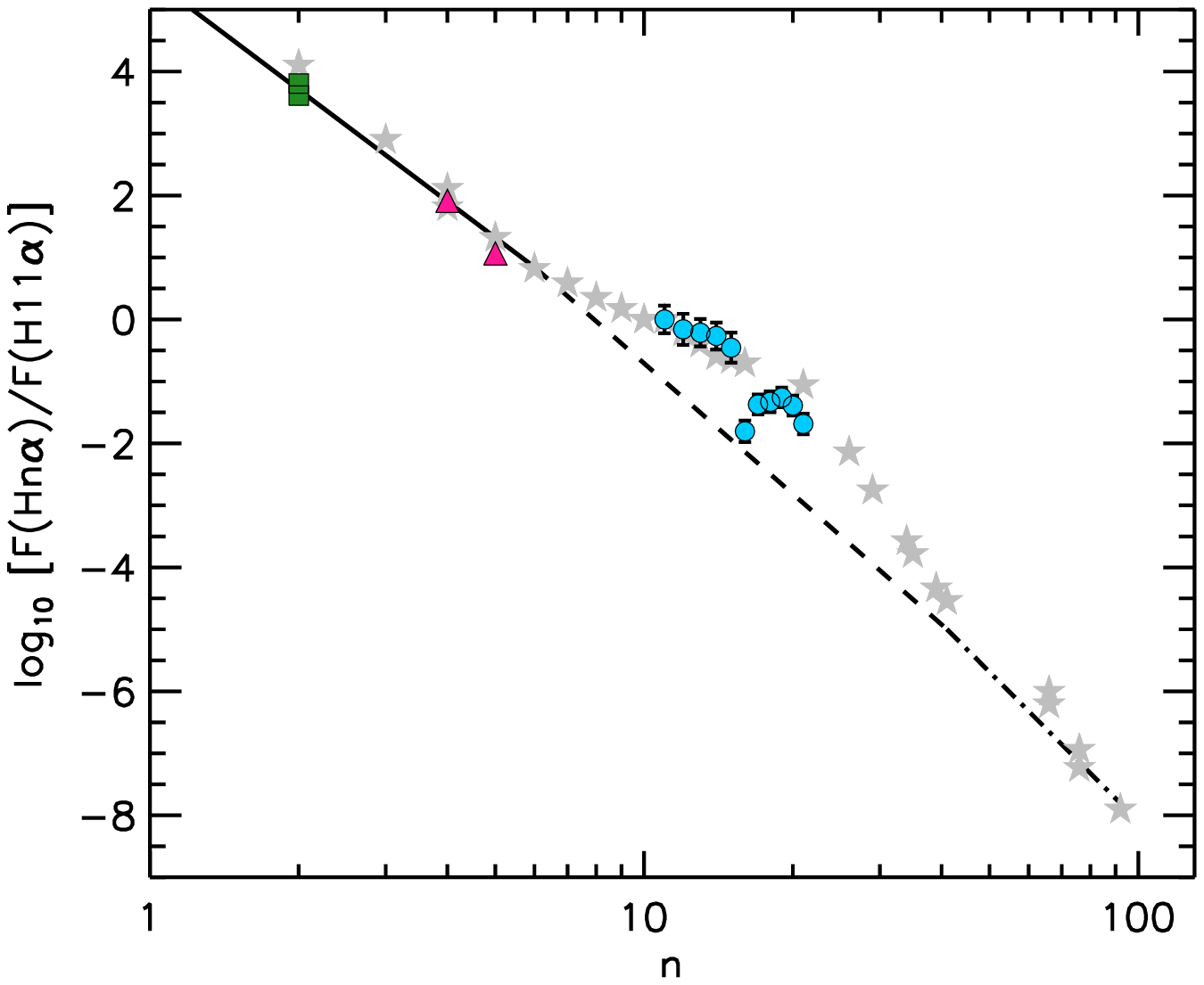}
   \caption{Hn$\alpha$ lines in the Mz~3 FIR/submm spectrum. Mz~3 \textit{Herschel} observations are represented by dots (this work). PACS fluxes are the values integrated in the three central spaxels. SPIRE fluxes are from the central pointing in the top panel and for the lobe pointing in the bottom panel (central bolometer only for both cases). In both panels, ISO observations of Br$\alpha$ and Pf$\alpha$ are represented by triangles \citep[][ this work]{2003ApJS..147..379S}, and H\,$\alpha$ observations of ESO 1.52 and CTIO 1.5m optical telescopes by squares \citep{2002MNRAS.337..499Z,2003MNRAS.342..383S}. Gray stars are observations of MWC~349A compiled by \citet{1998A&A...333L..63T}. The solid, dashed, and dot-dashed curves are the theoretical expectation for the nebular spontaneous emission for the optical to MIR, FIR to millimetre, and radio regimes as discussed in Sect. \ref{discuss}.}
\label{herschel_h_lines}
\end{figure}

\paragraph*{The spontaneous emission of an optically thin gas does not fit the Mz~3 HRL ratios.} The plots in Fig.~\ref{herschel_h_lines} show the \textit{Herschel} PACS and SPIRE HRL fluxes relative to H11$\alpha$ plotted as a function of $n$ for Mz~3 (dots). The top panel in Figure \ref{herschel_h_lines} uses the SPIRE centre pointing, while the bottom plot uses the southern lobe pointing.

As mentioned in the previous section, the SPIRE beam depends on the wavelength. Moreover the SPIRE beam is different from the PACS field of view (FOV). Also in the previous section, we note that most of the emission comes from the region covered by the three central spaxels [1,2], [2,2], and [3,2].  \citet{2002MNRAS.337..499Z} and \citet{2000MNRAS.312L..23R} showed, respectively, that most of the H\,$\beta$ emission (detected within a slit across the nebular major axis) and most of the H\,$\alpha$\footnote{Note that here we are using the regular way of representing the $\alpha$ and $\beta$ lines for the Balmer series, i.e. omitting the number 2 for the lower $n$ level of the series} emission are produced in a region within 15$\arcsec$ of the central source. Therefore it is natural to assume that the PACS three central spaxels region mentioned above and the SPIRE central bolometer for central pointing in any wavelength cover most of the HRL emission. Thus no beam corrections need to be applied for the comparison in Fig. \ref{herschel_h_lines} top. The case of SPIRE lobe pointing will be discussed later in this section.


In Fig. \ref{herschel_h_lines}, we also included ISO MIR\footnote{Our measurements were taken from the spectra obtained from the ISO SWS Atlas constructed by \citet{2003ApJS..147..379S}. For these lines, we used {\sc Splat/Starlink} to make the measurements.} and optical data. The ISO $\alpha$ lines included are  H4$\alpha$ (Bracket series) and H5$\alpha$ (Pfund) and the optical is H\,$\alpha$ (Balmer). Since the ISO FOV of 14$\arcsec \times$~20$\arcsec$ (which is centred on the central source with the large dimension along the main nebular axis) should also include most of the HRL emission, no correction of the fluxes to account for different beams/FOV is required. H\,$\alpha$ to H\,$\beta$ ratios have been reported by \citet{2002MNRAS.337..499Z} (H\,$\alpha$/H\,$\beta =$~3.26) and \citet{2003MNRAS.342..383S} (H\,$\alpha$/H\,$\beta =$~5.10). The difference between the ratios is due to different de-reddening methodologies. The uncorrected ratios obtained by each group are very similar despite using different instrumentation and FOV. To calculate the H\,$\alpha$ absolute flux, we assume the H\,$\beta$ flux estimated from radio emission at 6~cm by \citet{2005A&A...444..861P}, $F_{\rm H\,\beta} =$~2.86~$\times$~10$^{-10}$~erg~cm$^{-2}$~s$^{-1}$. The gas at such frequency seems to be optically thin, while there is evidence that the direct measure of H\,$\beta$ is probing only the outer gas layers \citet{2005A&A...444..861P}. This is important to consider as the FIR lines probe regions deep into the dense core, as we will discuss below. Although the H\,$\beta$ flux includes the emission from the whole nebula, the H\,$\alpha$ to H\,$\beta$ ratios we used only probe the emission within slits widths from 1.5$\arcsec$ to 2.0$\arcsec$. The H\,$\alpha$ to H\,$\beta$ ratios should however be representative, since the slits cover regions where the emission significantly contributes to the total emission.


The optical to MIR HRL emission of regular PNe and H~{\sc ii} Regions can usually be well fitted by the calculated emission for an optically thin gas dominated by spontaneous emission \citep{2006agna.book.....O}. We attempted to fit the Mz~3 observed HRL ratios curve using the emissivities provided by \citet{1995MNRAS.272...41S}. We explored coefficients for the whole range of physical conditions (temperature and density) provided by the authors, but found no reasonable curve that simultaneously fits all the observations (optical to submm).


\paragraph*{Mz~3 and MWC~349A HRLs flux ratios are similar.}
In Fig. \ref{herschel_h_lines}, we include the MWC~349A HRL ratios \citep{1998A&A...333L..63T} for comparison. MWC~349A is a massive B[e]SG star and the first object where HRL laser has been detected \citep{1989A&A...215L..13M}. As in Mz~3, MWC~349A has a circumstellar bipolar nebula produced by intense outflows and a dense disc seen nearly edge-on in its compact central core \citep{2001ApJ...562..440D}. 

Although the Mz~3 data are not as comprehensive as for MWC~349A in terms of $n$ coverage, we can see that the behaviour of the ratios of both objects are very similar.


\citet{1996ApJ...470.1134S} and \citet{1998A&A...333L..63T} studied the behaviour of the H$n\alpha$ as a function of $n$ for MWC~349A. For $n <$ 6-7 (optical to MIR), the curve is steeper than the curve expected for the spontaneous emission of an optically thin gas, which indicates that the lines are at least partially optically thick \citep[see][and references therein]{1998A&A...333L..63T}. A curve with a $n^{-6}$ dependence is expected in this case \citep{1996ApJ...470.1134S}. In the radio regime ($n >$ 40), the free-free opacity is important and the HRL ratios must follow a $n^{-8}$ behaviour \citep[assuming a spherically symmetric, constant velocity outflow at $T_{\rm e} =$~10$^4$~K;][]{1996ApJ...470.1134S}. The theoretical curves in Fig. \ref{herschel_h_lines} representing the ratios in both regimes is connected by a curve with a $n^{-7}$ dependence following \citet{1996ApJ...470.1134S}. It is clear from Fig. \ref{herschel_h_lines} that the data points for MZ 3 follow those of MWC 349A. 

In the optical to MIR regime and the radio regime, the HRL emission of MWC~349A follows the solid curve. The MWC~349A FIR to mm lines, on the other hand, clearly deviate from the solid curve. The extensive dataset of MWC~349A allowed \citet{1996ApJ...470.1134S} to show that such enhancement in the emission for $n$ in the interval 8 to 40 is due to amplification by stimulated emission (laser effect) occurring in its core \citep[see also][]{1998A&A...333L..63T,2013A&A...553A..45B}. The HRL laser emission detected in MWC~349A is believed to be produced in the ionized surface layer of its evaporating disc \citep{1994A&A...283..582T,2014A&A...571L...4B}. MWC~349A has been extensively studied and the action of laser effect \textbf{on} its HRLs is well established.


\paragraph*{The laser emission is produced in the core of Mz~3.}
In the previous section, we show that the HRL emission detected in the PACS observations of Mz~3 is concentrated mostly in the central spaxel. Similar evidence is found in the SPIRE lobe pointing observations, as we will discuss in the following. 

In the bottom panel of Fig. \ref{herschel_h_lines}, we provide HRL ratios derived from the SPIRE central bolometer spectrum of the southern lobe instead of the SPIRE central pointing. Although the comparison with the PACS three-spaxels central region is not strictly correct as we are not probing the same region, it is informative and supports the idea that the laser effect is produced in the core and not in the lobes. As for the case of the top panel in Fig. \ref{herschel_h_lines}, no correction for the different beam sizes was made. In the figure, we can see that the SPIRE line ratios decrease towards the ratios not influenced by laser effect (solid curve) for low $n$ values. This behaviour is due to the decrease of the SPIRE beam size at shorter wavelengths (i.e. for lower $n$ values; see Fig. \ref{spire_lines_beam}). As the beam size decreases, the core contribution to the emission inside the beam also decreases, and the line ratios approach the `non-laser' values. This behaviour is consistent with the scenario above where the core region is responsible for the laser emission.

\paragraph*{Mz~3 and MWC~349A core conditions are similar.} 


MWC~349A has been extensively studied in the literature. Its evolutionary fase is, however, still under debate. 
This B[e] star shows an ionized bipolar structure identified on scales from 0.05 to 1.5 pc as shown by \citet{2004ApJ...610..827T} and \citet{2012A&A...541A...7G}. In the waist of this bipolar nebula, there is a very compact core comprised of a a central source that ionizes the surface of the surrounding disc. 
The central source is likely a close binary system \citep{2012A&A...541A...7G}. The secondary appears to be a low-mass star and the primary a massive B0 or late O star with stellar parameters in the following ranges: $T_{\rm eff} =$~20--30~kK, $L_{\star}=$~3$\times$10$^4$--8$\times$10$^5$~$L_{\sun}$, $M =$~30--40~$M_{\sun}$ \citep[][and references therein]{2002A&A...395..891H,2012A&A...541A...7G}. 
The bipolar structure is formed by the strong winds produced at the rotating disc surface at radii smaller than 24~AU \citep{2014A&A...571L...4B}. The outflows have velocities of a few tens of km~s$^{-1}$ \citep{2002IAUS..206..226M,2014A&A...571L...4B}. The inclination of the bipolar structure with respect to the plane-of-sky is small ($\sim$15$\degr$), which indicates that the disc is also seen almost edge-on \citep{1994ApJ...428..324R}. 
The best model for MWC~349A found by \citet{2013A&A...553A..45B} yields an inclination of 8$\degr$ with respect to the line-of-sight for the disc. 
The size of the disc is estimated to be roughly between 50 to 120 AU \citep{2017ApJ...844...22S}. The inner radius should be less than 3 AU \citep{2013A&A...553A..45B}.


According to \citet{1994A&A...283..582T}, the characteristics of the submm lines \textit{``can be explained if the masers are on the ionized surface of a rotating disc with an additional small velocity component direct toward the center of rotation''}. The fact that the disc is seen edge-on is significant, as this will maximise the laser amplification length along the emitting surface. It is only necessary that the disc is seen roughly edge-on to have a sufficient effect \citep{1998A&A...333L..63T}. In MWC~349A, laser effect has been detected in HRLs with $n$ between 10 and 40 \citep[MIR to radio;][]{1996ApJ...470.1134S,1998A&A...333L..63T,2002IAUS..206..226M,2013A&A...553A..45B}. Calculations by \citet{2000A&A...361.1169H} and Calculations by \citet{2000A&A...361.1169H} and \citet{1998A&A...333L..63T} indicated that the densities in the zones where the laser lines are produced are $\sim$10$^8$~cm$^{-3}$.


The structure and physical conditions of the core of Mz~3 is very similar to those of MWC~349A. Mz~3 exhibits a bipolar structure with a very compact core where a nearly edge-on disc surrounds the ionizing central source. 
\citet{2004A&A...426..185S} inferred that the lobes of Mz~3 are inclined in $\sim$17$\degr$ in relation to the plane-of-sky. Outflows velocities between 130 and 500~km~s$^{-1}$ have been observed \citep{2004A&A...426..185S, 2000MNRAS.312L..23R}. 

Photoionization analysis indicates that the ionizing source of Mz~3 has $T_{\rm eff} =$~39500~K and $L_{\star} =$~2300--10000~$L_{\sun}$ \citep{2010A&A...517A..95P,2003MNRAS.342..383S}. 
However, the presence of a large flux of hard X-Rays (E > 1keV) in the core seems to indicate that the central source may be even more luminous, but deeply embedded by X-ray absorbing material \citep{2003ApJ...591L..37K}. This is in line with the high extinction estimated for the Mz~3 core previously mentioned. 

Using optical iron diagnostic line ratios,  \citet{2002MNRAS.337..499Z} determined that the iron emission observed from the core of Mz~3 is emitted by a region with densities around 10$^{6.5}$~cm$^{-3}$.  \citet{2003MNRAS.342..383S} used the observed suppression of a few atomic forbidden lines in the core and the concept of critical densities to estimate that the core region must have densities in excess of 10$^{6}$-10$^{7}$~cm$^{-3}$.  In both cases, the authors made clear that regions with densities above the forbidden lines critical densities may still exist, but can not be probed by such lines.
\citet{2007A&A...473L..29C} obtained a model for the Mz~3 disc, which was recently improved by \citet{2017arXiv170500120M}.  According to those works, the disc is rather flat, seen nearly edge-on and rich in amorphous silicates. The dust mass in the core is 2~$\times$~10$^{-5}$~$M_{\sun}$ \citep{2010A&A...514A..54G}. With a inner and outer disk radius of $\sim$10~AU and $\sim$250~AU, respectively, and a disk thickness on the order of 10 AU, an average density of 2~$\times$~10$^{8}$~cm$^{-3}$ is obtained if a dust-to-gas ratio of 100 is assumed. This is a rough approximation, but indicates a core density similar to that at MWC~349A and a density necessary for the production of FIR/submm laser lines exist.


\citet{2012ApJ...754...28Z} found many similarities between the Mz~3 and NGC~2392 structures and abundances, which made the authors propose that both PNe have a similar origin in a binary system where a giant star is responsible for the central outflows being ejected inside the planetary nebula previously produced by its now compact companion. 
NGC~2392 has a structure very similar to Mz~3, but it is seen almost pole-on, which can explain the absence of H~{\sc i}~laser lines in the NGC~2392 \textit{Herschel} spectrum. Furthermore, \citet{2012ApJ...754...28Z} have not found evidence of very high density regions in the core of NGC~2392.


We cannot determine the exact extent of the $n$ range where laser is acting in Mz~3 due to the lack measurements at MIR, short wavelength FIR, and mm ranges (Fig. \ref{herschel_h_lines}). However, if the low-$n$ onset of laser amplification is close to $n =$~8--10 as for MWC~349A, which seems reasonable from Fig. \ref{herschel_h_lines}, the density of the emitting gas could reach 10$^{8}$--10$^{10}$~cm$^{-3}$ in the Mz~3 core \citep{1996ApJ...470.1118S}. The existence of densities exceeding 10$^6$~cm$^{-3}$ were determined by for Mz~3 \citet{2003MNRAS.342..383S} from optical emission.


Considering the similarities in the HRL emission of Mz~3 and MWC~349A and in the relevant physical structure and conditions of their cores as discussed above, it is natural to expect that laser effect may take place in the Mz~3 core. It is then reasonable to conclude that laser effect is a likely explanation to the detection and the observed enhancement in the FIR/submm HRL intensity ratios we observed in the core of Mz~3 with \textit{Herschel}.


The discovery of laser effect acting on the FIR/submm HRL emission puts Mz~3 on a very short list of objects where HRL lasers have been detected. Table \ref{laser_objects} lists those objects and some of their properties. Besides MWC~349, it has been shown that HRL lasers occur in MWC~922, $\eta$ Carinae, Cepheus A HW2, MonR2-IRS2, and M~82. The list includes objects of very different classes: one PN/Symbiotic, two B[e] stars, a blue luminous variable, two young stellar objects, and a galaxy. However, all the objects have in common a strong ionized bipolar outflow and a dense disc structure associated with the HRL laser-emitting region.


\begin{table*}
\centering
\caption{Properties of Objects with H Laser Emission}
\label{laser_objects}
\begin{tabular}{cccccccc}
\hline
Object	&	Ref. Detection	&	Classification I $^a$	&	Classification II$^b$	&	Strong Winds/	&	Outflow	&	Disc	&	Ref.$^c$	\\
	&	of H laser lines	&		&		&	 Outflows	&	Morphology	&		&		\\
\hline             
Mz~3 & 1 & Planetary Nebula & Young PN & Yes & Bipolar & Yes & 2--8 \\
 &  &  & pre-PN &  &  &  &  \\
 &  &  & Symbiotic &  &  &  &  \\
MWC~349A & 9 & Emission-line Star & B[e]SG star & Yes & Bipolar & Yes & 10--14 \\
 & & & Luminous Blue Variable &  &  &  &  \\
MWC~922 & 15 & Emission-line Star & B[e] FS CMa star & Yes & Bipolar & Yes & 15--16 \\
$\eta$~Car & 17 & Emission-line Star & Luminous Blue Variable & Yes & Bipolar & Yes & 17--19 \\
Cep~A~HW2 & 20 & Young Stellar Object & -- & Yes & Bipolar & Yes & 20--22 \\
MonR2-IRS2 & 23 & Star in Cluster & Compact YSO & Yes & Bipolar & Yes & 23--24 \\
 &  &  & TTauri Star &  &  &  &  \\
M82 & 25 & Interacting Galaxies & Starburst Galaxy & Yes & Bipolar & Yes & 25--26 \\

\hline      

\multicolumn{8}{l}{Notes:$^a$ Classification from CDS/Simbad. $^b$ Classification from other sources and complementing data (see references in the last}\\

\multicolumn{8}{l}{column). $^c$ References for data in columns 4 to 7.}\\

\multicolumn{8}{l}{References: 
(1) This Work;
(2) \citet{1978ApJ...221..151C}; 
(3) \citet{1983MNRAS.204..203L}; 
(4) \citet{2001A&A...377L..18S}; 
}\\
\multicolumn{8}{l}{ 
(5) \citet{2004MNRAS.354..549B}; 
(6) \citet{2011MNRAS.413..514C}; 
(7) \citet{1985MNRAS.215..761M}; 
(8) \citet{2007A&A...473L..29C}; 
}\\
\multicolumn{8}{l}{ 
(9) \citet{1989A&A...215L..13M}; 
(10) \citet{2016MNRAS.456.1424A}; 
(11) \citet{1985ApJ...297..677W}; 
(12) \citet{2012A&A...541A...7G}; 
}\\
\multicolumn{8}{l}{ 
(13) \citet{1994A&A...283..582T}; 
(14) \citet{2011A&A...530L..15M}; 
(15) \citet{2017A&A...603A..67S}; 
(16) \citet{2017A&A...601A..69W}; 
}\\
\multicolumn{8}{l}{ 
(17) \citet{1995A&A...295L..39C}; 
(18) \citet{1993A&A...268..283M}; 
(19) \citet{2014ApJ...791...95A}; 
(20) \citet{2011ApJ...732L..27J}; 
}\\
\multicolumn{8}{l}{ 
(21) \citet{2007ApJ...661L.187J}; 
(22) \citet{2014A&A...562A..82S}; 
(23) \citet{2013ApJ...764L...4J}; 
(24) \citet{2008hsf1.book..899C}; 
}\\
\multicolumn{8}{l}{ 
(25) \citet{1996ApJ...465..691S}; 
(26) \citet{2004ApJ...616..783R} 
}\\
\end{tabular}
\end{table*}

\section{Conclusions} \label{conclude}

In this paper, we report the detection of hydrogen recombination laser lines in the FIR to submm spectrum of Mz~3 observed with the \textit{Herschel} PACS and SPIRE instruments. Comparison of optical to submm HRL lines to theoretical calculations indicates that there is an enhancement in the FIR to submm HRLs, which explains their unexpected detection. The likely explanation for this enhancement is the occurrence of laser effect. Laser effect offers a natural explanation since: 

\begin{enumerate}

\item the available Mz~3 optical to submm HRL $\alpha$ line intensity ratios are not well reproduced by the spontaneous emission of an optically thin ionized gas, which is typical for the nebular gas in planetary nebula; 

\item the compact core is responsible for a large fraction of the Mz~3 \textit{Herschel} HRLs emission;

\item the line intensity ratios for Mz~3 are very similar to those in the core emission of the notorious star MWC~349A, where it is well established that laser effect is responsible for the enhancement of HRLs in the \textit{Herschel} wavelength range; 

\item the physical characteristics in the core MWC~349A that are responsible for producing the conditions for laser effect, i.e. dense equatorial disc seen nearly edge-on and intense ionized outflows, are also present in the Mz~3 core. 

\end{enumerate}

Our comparison of observations to models of laser emission from the literature indicates the presence of a dense and ionized gas ($n_{\rm H}$~>~10$^8$ cm$^{-3}$) in the core of Mz~3. For the surrounding lobes, the empirical analysis of forbidden lines indicates densities around 4500~cm$^{-3}$.

We have presented compelling evidence to support that laser effect is acting on the HRLs in Mz~3. Future work will improve the coverage of HRLs from Mz~3 to complete the curve shown in Fig.~\ref{herschel_h_lines}, as well as, resolved line velocity profiles. Velocity profiles can be used to constrain detailed models of the core structure of Mz~3. Sub-arcsecond submm ALMA observations would be ideal to isolate the central region from the rest of the nebula and resolve the laser components.

\section*{Acknowledgements}
 
I.A. acknowledges partial financial support of CNPq/Brazil. Studies of interstellar chemistry at Leiden Observatory are supported through the advanced-ERC grant 246976 from the European Research Council, through a grant by the Dutch Science Agency, NWO, as part of the Dutch Astrochemistry Network, and through the Spinoza prize from the Dutch Science Agency, NWO. S.W. acknowledges the Leiden/ESA Astrophysics Program for Summer Students (LEAPS), which supported his visit to the Leiden Observatory. P.v.H. is supported by the Belgian Federal Science Policy Office under contract No. BR/143/A2/BRASS. MO was supported by the research fund 106-2811-M-001-119 and 104-2112-M-001-041-MY3 from the Ministry of Science and Technology (MOST), R.O.C. We thank S. Akras for useful discussions about Mz~3. This research has made use of NASA's Astrophysics Data System and the SIMBAD database, operated at CDS, Strasbourg, France. Some of the data presented in this paper were obtained from the Mikulski Archive for Space Telescopes (MAST). STScI is operated by the Association of Universities for Research in Astronomy, Inc., under NASA contract NAS5-26555. Support for MAST for non-HST data is provided by the NASA Office of Space Science via grant NNX09AF08G and by other grants and contracts.


\bibliographystyle{mnras}  
\bibliography{references}      






\bsp	
\label{lastpage}
\end{document}